\documentclass[aps,reprint,onecolumn,floatfix,nofootinbib,12pt]{revtex4-2}
\usepackage{hyperref}
\hypersetup{
  breaklinks=true, 
  colorlinks=true, 
  pdfusetitle=true, 
}
\usepackage{float}
\usepackage{epsfig}
\usepackage{graphicx}
\usepackage{amsmath}
\usepackage{amsfonts}
\usepackage{amssymb}
\usepackage{bm}
\usepackage{bbold}
\usepackage{epsfig}
\usepackage{graphicx}
\usepackage{color}
\usepackage{pictex}
\usepackage{mathtools}
\usepackage{extarrows}
\usepackage{footnote}
\usepackage{footmisc}
\usepackage{setspace}
\newcommand{\s}{\scriptscriptstyle}
\begin{document}
\title{Fermi surface as a quantum critical manifold: gaplessness, order parameter, and scaling in $d$-dimensions}
\author{Gennady Y. Chitov}
\affiliation{Bogoliubov Laboratory of Theoretical Physics,
Joint Institute for Nuclear Research, Dubna 141980, Russia}
\date{\today}

%
%
\begin{abstract}
We study several models of $d$-dimensional fermions ($d=1,2,3$) with an emphasis on the properties of their gapless (metallic) phase. It occurs at $T = 0$ as a continuous transition when zeros of the partition function reach the real range of parameters.
Those zeros define the $(d-1)$-manifold of quantum criticality (Fermi surface). Its appearance or restructuring correspond to the Lifshitz transition.
Such $(d-1)$-membrane breaks the symmetry of the momentum space, leading to gapless excitations, a hallmark of metallic phase.
To probe quantitatively the gapless phase we introduce the geometric order parameter as $d$-volume of the Fermi sea. From analysis of the chain, ladder, and free fermions with different spectra, this proposal is shown to be consistent with scaling near the Lifshitz points of other quantities: correlation length, oscillation wavelength, susceptibilities, and entanglement. All the (hyper)scaling relations are satisfied. Two interacting cases of the Tomonaga-Luttinger ($d=1$) and the Fermi ($d=2,3$) liquids are analysed, yielding the same universality classes as free fermions.
\end{abstract}
\maketitle

%
%
%
\section{Introduction}\label{Intro}
%
%
%
%

The initial motivation for this work was in the effort to better understand the physics of the field-induced transition between the gapped and incommensurate
(IC) gapless phases in low-dimensional spin systems.
Quantum critical properties of chains and ladders in magnetic field have been actively studied over the years.
(See, e.g., \cite{Affleck:1997,Totsuka:1997,ChitraGiam:1997,Cabra:1999,*Cabra:2000,Delgado:2000,Japaridze:2006,Japaridze:2009,Li:2013,Hatsugai:2015,Mahdavifar:2017M,Ding:2020,Toplal:2022}).
The key property is the existence of quantized plateaux in the unique magnetization curve, corresponding to the gapped incompressible phases, continuously connected by the curves of arbitrary real magnetization. The latter correspond to the gapless incommensurate (IC) phase(s).
The necessary quantization condition for appearance of such plateaux in generic chains or ladders was derived in \cite{Affleck:1997}.
We will call it the \emph{plateau theorem} which states:\\
\emph{The magnetization plateau with rational (fractional) magnetization and a finite excitation gap is only possible if condition
\begin{equation}
 P(s- \mathcal{M}) \in \mathbb{Z}
\label{plateaueq}
\end{equation}
is satisfied.} Here $P$, $s$, and $\mathcal{M}$ are periodicity, spin, and magnetization, respectively.
Periodicity $P$ is defined as the number of spins in the unit cell in the ground state.
This theorem is a generalization of the Lieb-Schultz-Mattis (LSM) theorem \cite{LiebSM:1961} for the case of magnetic field.
The plateau theorem generalized for a quantum many-particle fermionic system defined on a periodic lattice with an exactly conserved particle number, states that a finite excitation gap is possible only if the particle number per unit cell of the ground state is an integer, i.e., at commensurate (rational) filling  \cite{Oshikawa:2000}.
Since the spin-$\frac12$ and spinless fermionic models can be mapped onto each other by a judiciously chosen Jordan-Wigner transformation,
one can always make a correspondence between the results for spins or fermions. In terms of fermions, the transitions between the IC gapless and the incompressible gapped (plateau) phases, are the metal-insulator transitions.

We want to pay a special attention to the gapless IC (metallic) phase. At first sight, no particular order parameter can be
attributed to it, except of a standard notion of the algebraic order, meaning that all the correlation functions have power-law decay in this phase.
Recently the Lifshitz transitions have been reported in the IC gapless phases of several spin and fermionic models
\cite{Cazalilla:2022,Alvarez:2022,BeradzeNerses:2023}. These transitions are related to the change  of number of the Fermi points in chains or ladders,
and they do not contradict the plateau theorem, as far as the system remains gapless. Restructuring of the Fermi surface (points in one spatial dimensions)
is the Lifshitz transition  \cite{Lifshitz:1960}.

The question we want to address in this work: what could be said beyond the statement that a change of the topology of the Fermi surface (FS) is a phase transition? Can we quantify its measure, and even to attribute a certain order parameter to the mere existence of such a surface, beyond counting the Fermi points and/or calculating the topological winding numbers?
The appearance of the FS constitutes a quantum phase transition by itself, as has being emphasized by Volovik for some time \cite{Volovik:2003,Volovik:2007,*Volovik:2017}.

Probably the most fundamental and unbiased rigorous approach to detect a phase transition is to analyse zeros (or poles) of the partition function $\mathcal{Z}$.  This approach, pioneered by Yang and Lee \cite{YangLee:1952,*LeeYang:1952}, was proven to be very successful for studies of various models at equilibrium and even out of it. For a short list of references, see, e.g., \cite{Fisher:1965,Fisher:1980,Matveev:2008,Bena:2005,TongLiu:2006,Wei:2014,Chitov:2021,Chitov:2022DL}, and more references in there.
From exact calculations for several $d$-dimensional free fermionic models ($d=1,2,3$) we show that a phase transition can occur only at $T = 0$, signalled by zeros of the partition function in the real range of parameters. Those Lee-Yang zeros are shown to coincide with zeros of the inverse Matsubara single-particle Green's function, since
$\mathcal{Z} \propto \mathrm{det} \hat G^{-1}$. The manifold of those zeros at $T=0$ defines the $(d-1)$-dimensional FS which separates the occupied and empty fermionic states.
Thus the Fermi energy defines the surface of quantum criticality (the manifold of Lee-Yang zeros), and its restructuring  or even an appearance constitutes a quantum phase transition. See also  \cite{Volovik:2003,Volovik:2007,*Volovik:2017} and  \cite{Horava:2005}. Note that relating the FS to zeros of the inverse
Green's function is just a standard working definition (see, e.g., \cite{AGD:1963}), while identifying FS as a manifold of quantum criticality is a nontrivial rigorous result (see also \cite{Chitov:2021}) following from the fact that it is a manifold of the Lee-Yang zeros in the limit $T\to 0$.
Appearance of a $(d-1)$-dimensional membrane in the momentum space \cite{Horava:2005} breaks the initial symmetry of the latter, and as a consequence gapless excitations around the FS appear as well, which is a hallmark of metallic phase.

To elaborate the above points and to put the description of this quantum transition in the quantitative framework, we introduce in this paper the order parameter $\mathcal{P}$ for the gapless metallic phase as the measure ($d$-dimensional volume) of the Fermi sea in the momentum space. We demonstrate that this proposal is consistent with the scaling  properties of other physical parameters, like the correlation length, the wavelength of the IC oscillations, the compressibility, and the entanglement. In particular, all the scaling and hyperscaling relations for corresponding critical indices, known from the theory of quantum phase transitions \cite{Imada:1998,Sachdev:2011}, are shown to be satisfied.
From analysis of two paradigmatic examples of interacting metallic states -- the Fermi and the Tomonaga-Luttinger liquids, we show that the interactions do not alter the essence of those results, and the universality class of those liquids is as of the free Fermi gas.

The rest of the paper is organized as follows: In Sec.~\ref{LadMod} we introduce the exactly-solvable fermionic  two-leg ladder model and study its critical properties. In Sec.~\ref{Gen} we propose the order parameter for a generic gapless (metallic) phase. The consistency of the scaling, order parameters, and the relations between critical indices are verified for the ladder model and $d$-dimensional fermions with different spectra. The Fermi and the Tomonaga-Luttinger liquids are also analysed in that section. Two Appendices contain derivations and most of analytical results for the ladder model and $d$-dimensional fermions with non-relativistic and Dirac spectra. For a smoother reading it is probably better to go through the Appendices after the Introduction. The results are summarized and discussed in the concluding Sec.~\ref{Concl}.

%
%
%
\section{Lifshitz transitions in two-leg ladder}\label{LadMod}
%
%
%
%

%
%
%
\subsection{Two-leg ladder model and plateaux}\label{Model}
%
%
%
%

\begin{figure}[h]
\centering{\includegraphics[width=6.0cm]{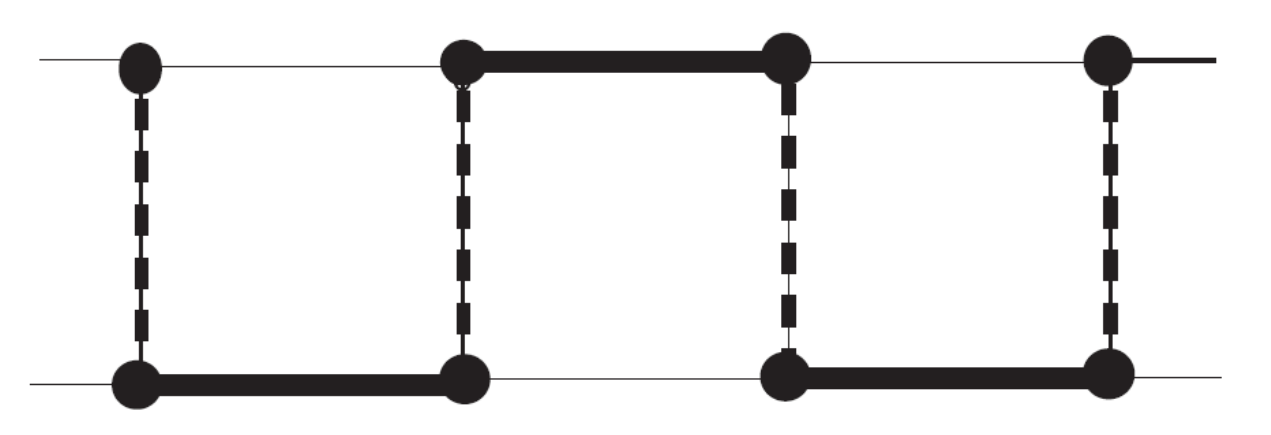}}
\caption{Two-leg ladder with staggered dimerization. Bold/thin/dashed lines represent, respectively, the stronter/weaker in-chain $J(1\pm\delta)$ and rung $J_\perp$ couplings of the spin Hamiltonian \eqref{Ham}. For the fermionic Hamiltonian  \eqref{Hmf} those lines represent corresponding hopping parameters.}
\label{SDLadder}
\end{figure}

The incentive model is the Heisenberg spin-$\frac12$ two-leg ladder in transverse magnetic field defined by Eq.~\eqref{Ham}.
The spin ladder \eqref{Ham} can be fermionized and mapped exactly onto an interacting Hamiltonian of spinless fermions
(see Appendix \ref{AppLad} for discussion on this). The physics we want to explore is not interaction-driven. It is fully imbedded
in the quadratic effective fermionic Hamiltonian, which can be viewed as a ``light" solvable version of \eqref{Ham}:
\begin{eqnarray}
  H_{F} &=& \frac12 \sum_{n=1}^{N}\sum_{\alpha=1}^{2} \Big\{ (-1)^{\alpha-1}
  t_{\alpha }(n)c^{\dag}_{\alpha}(n) c_{\alpha}(n+1) +H.c.
  -2\mathfrak{h}_{\alpha }(n) \Big[ c^{\dag}_{\alpha}(n) c_{\alpha}(n)-\frac12 \Big] \Big\} \nonumber \\
  &+&  \sum_{n=1}^{N} t_{\perp }c^{\dag}_{1}(n)c_{2}(n)+H.c  ~.
\label{Hmf}
\end{eqnarray}
Here
\begin{eqnarray}
\label{talpha}
  t_{\alpha }(n)= t [1+(-1)^{n+\alpha}\delta]~,\\
  \mathfrak{h}_{\alpha }(n)=\mu+(-1)^{n+\alpha}\mu_a~.
\label{halpha}
\end{eqnarray}
So we focus on the exactly solvable model of two-leg ladder with the tight-binding Hamiltonian \eqref{Hmf}. Its parameters are: in-chain dimerized hopping $t_{\alpha }(n)$, transverse hopping $t_\perp$, chemical potential $\mu$, and  modulated external potential $\mu_a$. See Fig.~\ref{SDLadder}.
From the symmetry of Hamiltonians \eqref{Ham}  and \eqref{Hmf} it suffices to consider only nonnegative values $(h, \mu, h_a, \mu_a) \geq 0$.
We shall use the natural units $\hbar=k_B=1$. In the following all dimensional quantities will be expressed in units of the energy scale $t$.

The Jordan-Wigner spin-fermion mapping allows us to establish correspondence between the induced magnetization per site $\mathcal{M}$ and the average number of fermions per site (filling) $\bar{n}$ as
\begin{equation}
\label{MNudef}
 \mathcal{M} \equiv \frac{1}{4N} \sum_{\alpha=1}^{2} \sum_{n=1}^{N}\langle\sigma_{\alpha}^{z}(n)\rangle~~\Longleftrightarrow~~
 \bar{n} - \frac12 \equiv  \frac{1}{2N} \sum_{\alpha=1}^{2} \sum_{n=1}^{N} \Big(\langle  c^{\dag}_{\alpha}(n) c_{\alpha}(n)\rangle -\frac12\Big)
\end{equation}

For a dimerized chain or staggered two-leg ladder ($s=1/2$) in the uniform or staggered transverse field ($P=2$) the theorem \eqref{plateaueq} predicts occurrence of two plateaux at $\mathcal{M}=0$ and $1/2$, i.e., at $\bar n=1/2,1$, according to \cite{Oshikawa:2000}.

Diagonalization of the Hamiltonian (\ref{Hmf}) yields four eigenvalues:
\begin{equation}
  E_{\pm\pm}(k)= \pm \varepsilon_\pm (k)-\mu ,~~\mathrm{where}~~
  \varepsilon_\pm (k) \equiv \sqrt{\mu_{a}^2+\cos^2k+ \big(\delta\sin k \pm t_{\perp}\big)^2}~.
\label{spectra_ss1}
\end{equation}
The fermions have four bands $\pm \varepsilon_\pm (k)$ within reduced Brillouin zone (BZ) $k \in [-\pi/2, \pi/2]$, and the band filling is controlled by the value of $\mu$.

\begin{figure}[h]
\centering{\includegraphics[width=12.0 cm]{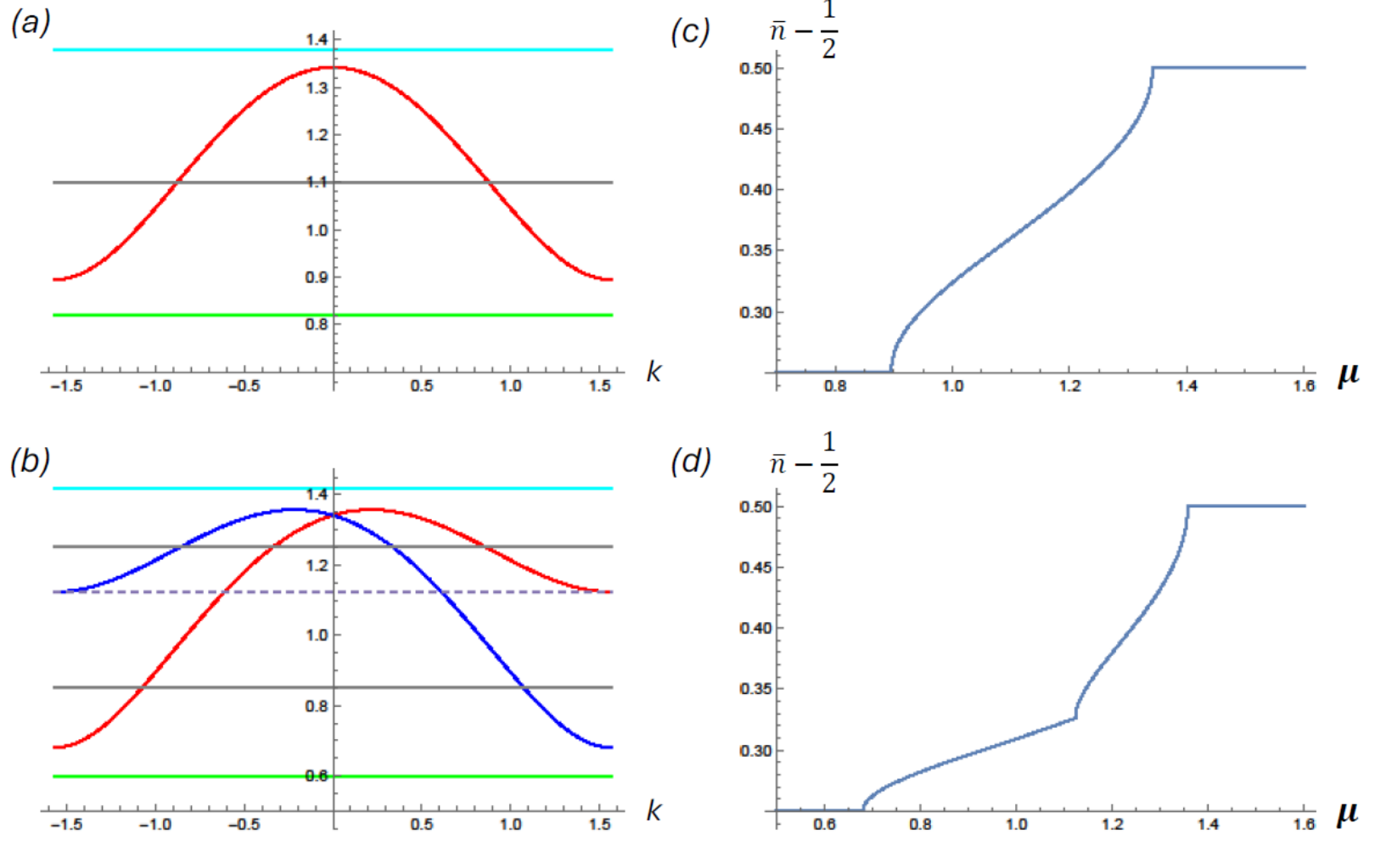}}
\caption{(a) Two identical bands $\varepsilon_+(k)=\varepsilon_-(k)$. Different values of $\mu$ (horizontal lines) correspond: empty band, $\mu<\mu_{c,1}$ (green); partially filled band from the BZ edges up to two Fermi points $\pm k_F$, $\mu_{c,1}<\mu<\mu_{c,2}$ (grey); totally filled band, $\mu> \mu_{c,2}$ (cyan).  (c)
The filling $\bar n$ for the spectrum shown in (a). Cf. \eqref{MNudef} for $\bar n -1/2 \leftrightarrow \mathcal{M} $ correspondence.
The plots (a,c)  are done for $\mu_a=0.4$, $\delta=0.0$, $t_\perp =0.8$.
(b) Two split bands $\varepsilon_+(k)$ and $\varepsilon_-(k)$ (shown for $\delta=0.25$, the same $\mu_a,t_\perp$).
In comparison to the case (a) the dashed line at $\mu=\mu_L$ and a grey line at $\mu_L<\mu<\mu_{c,2}$ (four Fermi points), are added.
(d) The filling $\bar n$ with a cusp at $\mu=\mu_L$ in the gapless phase.
Both cases (c) and (d) are in agreement with the plateau theorem \eqref{plateaueq}.}
\label{BandsPl}
\end{figure}

The filling can be readily found as\footnote{\label{LBs} Note that two lower bands $-\varepsilon_\pm(k)$ are always filled at $\mu \geq0$, their (constant) contribution is accounted for in \eqref{NuMan}, so we will not mention them in the following.  In particular, they are not shown in Fig.~\ref{BandsPl}.}
\begin{equation}
 \bar{n}=1-\frac{1}{4\pi}\int_{-\pi/2}^{\pi/2}\Big\{\mathrm{sign} \big(\varepsilon_+(k)-\mu \big)+ \mathrm{sign} \big(\varepsilon_-(k)-\mu \big) \Big\}dk~.
 \label{NuMan}
\end{equation}
In accord with the general theorem  (\ref{plateaueq}), the spectrum  \eqref{spectra_ss1} reveals two gapped phases with plateaux connected via quantum phase transitions of the second kind to a gapless incommensurate (IC) phase.

In the limiting case when two dimerized chains are decoupled ($t_\perp =0$) or the ladder is uniform  ($\delta =0$), the two bands \eqref{spectra_ss1} merge: $\varepsilon_+(k)=\varepsilon_-(k) \equiv \varepsilon(k)$. The evolution of $\bar n$ with $\mu$ through different phases can be readily inferred from Fig.~\ref{BandsPl}a.
The two plateaux at rational fillings $\bar{n}=\frac12, 1$ in the gapped phases are connected by a \textit{smooth} continuous curve of $\bar{n} \in \mathbb{R}$ in the IC gapless phase at $\mu_{c,1}<\mu<\mu_{c,2}$, see Fig.~\ref{BandsPl}c. The explicit formulas for this case follow as limits $\delta \to 0$ or $t_\perp \to 0$ from the equations of
Sec.~\ref{AppL}.

%
%
%
\subsection{Gapless-to-gapless Lifshitz transition in the IC phase}\label{Res}
%
%
%
%
A new feature occurs when two dimerized ($\delta \neq 0$) chains are coupled in a ladder ($t_\perp \neq 0$).
The band structure reveals a Lifshitz transition  within the gapless phase.
Two bands $\varepsilon_\pm(k)$  are split, even if the staggered potential $\mu_a$ is absent, see Fig.~\ref{BandsPl}b.
As a consequence, a new transition occurs at $\mu_L \in (\mu_{c,1},\mu_{c,2})$, when the number of Fermi points doubles $2 \rightarrow 4$, yielding a cusp of  $\bar{n}(\mu)$ in the IC phase, seen in Fig.~\ref{BandsPl}d. Explicit formulas for the critical values $\mu_\sharp$ are given by Eqs.(\ref{hc1},\ref{hL},\ref{hc2}). 
The full account of the analysis and exact results for the physical parameters are presented in Sec.~\ref{AppL}.

\begin{figure}[h]
\centering{\includegraphics[width=11.0 cm]{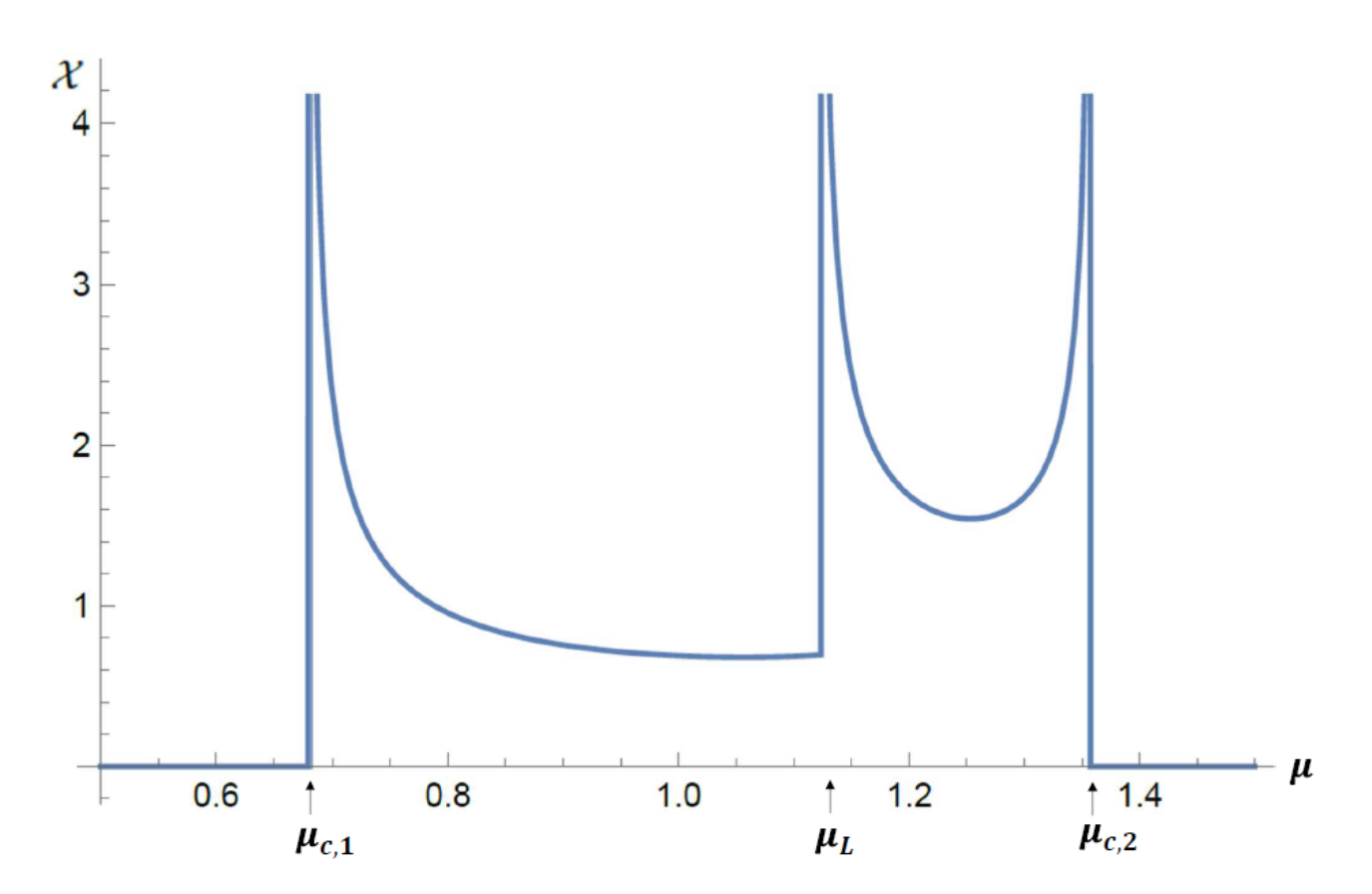}}
 \caption{Compressibility: the plot is done for $\mu_a=0.4$, $\delta=0.25$, $t_\perp =0.8$. For those parameters the critical points are: $\mu_{c,1} =0.68$, $\mu_L =1.12$, $\mu_{c,2}=1.36$.}
 \label{Comp}
\end{figure}

The compressibility is plotted in Fig.~\ref{Comp}.
Two plateaux in the gapped phases are the incompressible states of fermions, and the compressibility  is singular near the quantum critical points
$\mu_{c,1/2}$:
\begin{equation}
\label{chi12}
  \chi_\circ \propto \frac{1}{\sqrt{|\mu -\mu_{c,\sharp}|} }~,~~\mu \to \mu_{c,\sharp} \pm 0~,
\end{equation}
in agreement with earlier results \cite{Vekua:2009}.
The compressibility is also singular on one side from the Lifshitz transition in the gapless IC phase,
where $\bar{n}(\mu)$ demonstrates a cusp:
\begin{equation}
\label{chiL}
   \chi_\circ  \propto \frac{1}{\sqrt{\mu-\mu_L} }~,~~\mu \to \mu_L + 0~.
\end{equation}
So, for all three transitions the critical index of susceptibility $\gamma =1/2$.
With $\mu$ growing from zero and crossing over the upper critical value  $\mu_{c,2}$, the system undergoes a sequence of the second order transitions
accompanied by the change of the Fermi points number: $0 \rightarrow 2 \rightarrow 4 \rightarrow 2 \rightarrow 0$ (note that at $\mu=\mu_{c,2}$:
$k_{\s F,1} =- k_{ \s F,2}$). The Fermi wave numbers $k_{\s F,1/2}$ are given by Eq.\eqref{ypm}.
Two critical points  at $\mu_{c,1}$ and $\mu_{c,2}$ are the Lifshitz transitions as well, since the number of the Fermi points changes $0 \leftrightarrow 2$.
The transitions at $\mu_{c,1/2}$ are accompanied by the gap closure/openning at the boundary with the IC phase.
A particularly interesting feature of the Lifshitz transition at $\mu_L$ is that it occurs inside the quantum critical state, which is gapless and the correlation length is infinite. Regardless the presence or absence of the gap, changing of the Fermi points ($\pm 2$) is signalled by the same type of singularity, cf. Eqs.~\eqref{chi12} and \eqref{chiL}.

%
%
%
\subsection{Lee-Yang zeros and quantum transitions}\label{LYZsec}
%
%
%
To better understand the physical properties of the gapped phases and the gapless states on both sides from the Lifshitz points, we analyse zeros of the partition function of the model \eqref{Hmf} in the ground state. We follow the lines of the earlier work \cite{Chitov:2021,Chitov:2022DL}.
\begin{figure}[h]
\centering{\includegraphics[width=11.0 cm]{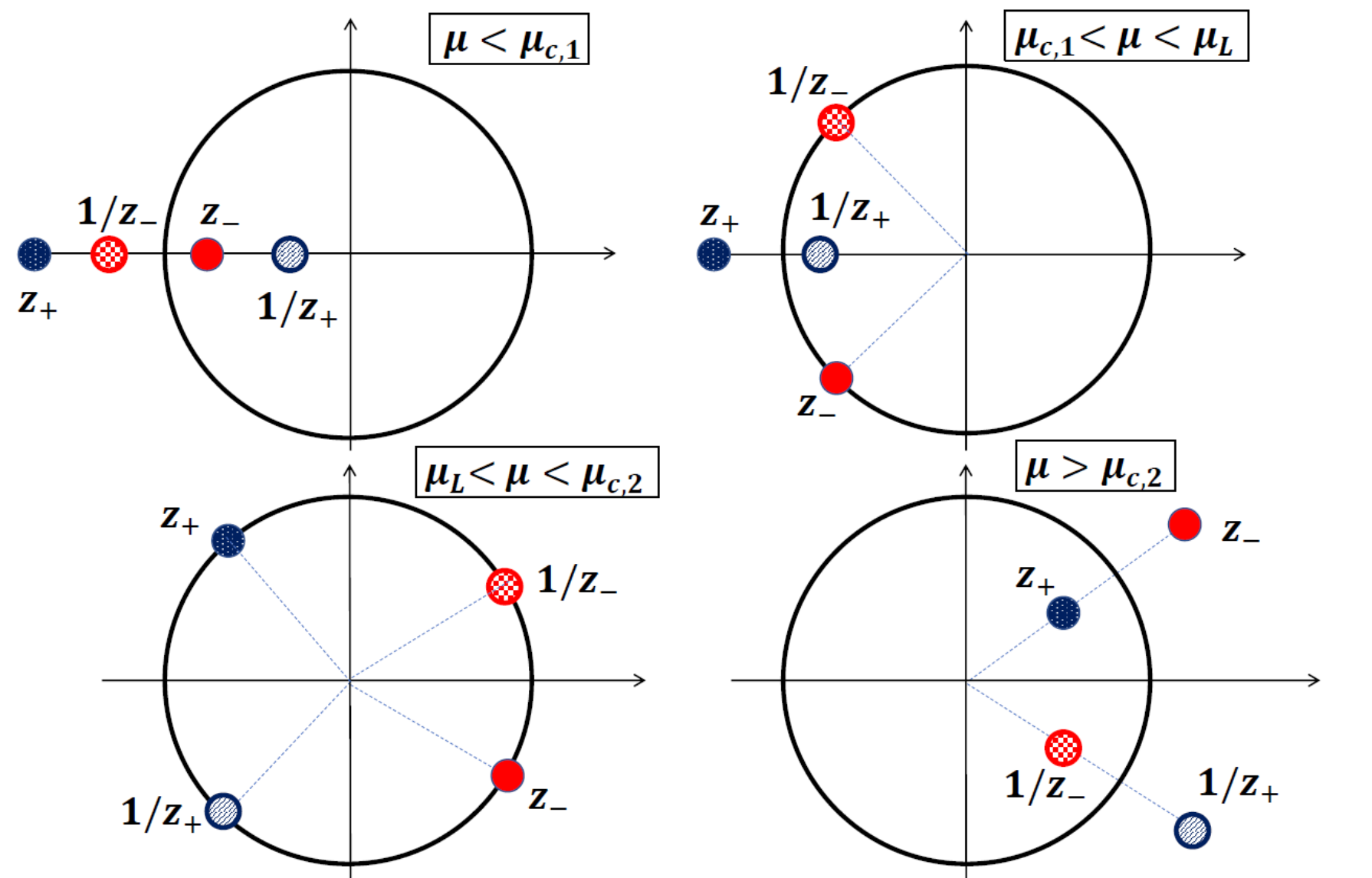}}
 \caption{Positions of the roots $z_\pm$ and $1/z_\pm$ on the complex plane with respect to the unit circle $|z|=1$, shown for four different phases.}
 \label{ZRoots}
\end{figure}
A complete analysis of zeros of the partition function of the model \eqref{Hmf} at $T=0$ and the physical implications of those results are presented in
Sec.~\ref{AppLYZL} of Appendix \ref{AppLad}. The key points are the following:

We parameterize the roots of the equation for zeros of the partition function by mapping the reduced BZ onto a unit circle on the complex plane
as $z \equiv e^{2i k}$.
It has four roots ($z_\pm,1/ z_\pm$), two of them  are mutually reciprocal (cf. \eqref{Z0zpm} and \eqref{ZpmY}).
The quantum critical points correspond to the physical roots when $k \in \mathbb{R}$, i.e., those which lie on a unit circle $|z_\pm|=1$.

Evolution of the positions of the roots on the complex plane with respect to the unit circle in four different phases is shown in Fig.~\ref{ZRoots}.

\begin{figure}[h]
\centering{\includegraphics[width=7.0 cm]{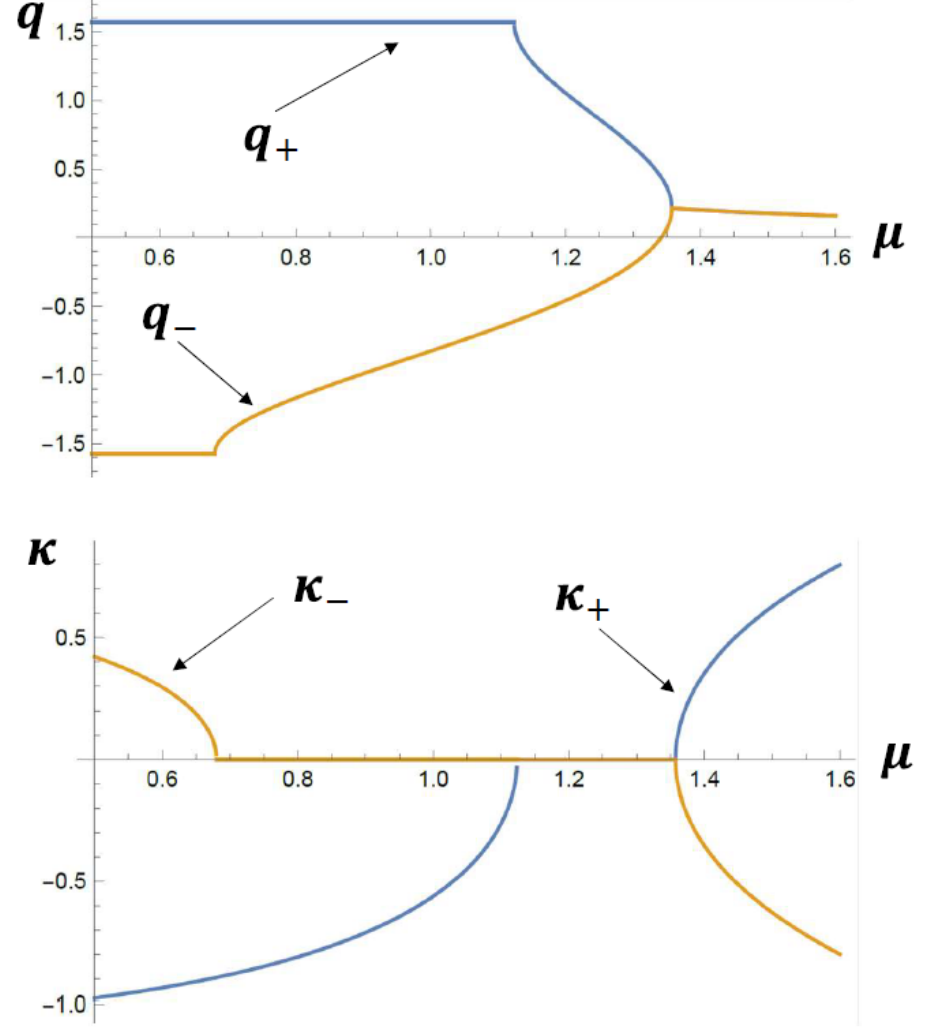}}
\caption{Real and imaginary parts of the complex wave number $k_\pm= q_\pm +i \kappa_\pm$, cf. Eqs.~\eqref{qkap} and \eqref{zk} in four different phases. The plot is done for $\mu_a=0.4$, $\delta=0.25$, $t_\perp =0.8$. For those parameters the critical points are: $\mu_{c,1} =0.68$, $\mu_L =1.12$, $\mu_{c,2}=1.36$.}
\label{KappaQ}
\end{figure}

(1). In the first gapped phase (plateau $\bar{n}=\frac12$) four roots \eqref{ZpmY} are real and negative; the couple $z_-$ and $1/z_-$ approaches the unit circle, merging on it at the quantum critical point $\mu =\mu_{c,1}$, where $z_-=-1$. \\
(2). The gap closes and the systems stays gapless in the IC phase, two roots are complex conjugate and move over the unit circle:
\begin{equation}
\label{Zpm1}
 \mu_{c,1}< \mu < \mu_L:~ z_-=e^{2 ik_{\s F,1}},~~1/z_-=z_-^\ast
\end{equation}
They determine the Fermi wave number $k_{\s F,1}$ of the IC oscillations.
Deeper in the gapless phase the other couple approaches the unit circle, merging on it at the Lifshitz critical point $\mu =\mu_L$ where $z_+=-1$. \\
(3). In the gapless phase after the Lifshitz transition all four roots evolve smoothly, staying on the unit circle
\begin{equation}
\label{Zpm2}
  \mu_L < \mu <\mu_{c,2}:~ z_\mp =e^{2 ik_{\s F,1/2}}~.
\end{equation}
The IC oscillations of correlation functions change their nature: now they are superposition of oscillations with two Fermi wave numbers $k_{\s F,1/2}$.
This is an observable consequence of the topological Lifshitz transition, when the Fermi sea changes its connectivity: the number of disconnected Fermi patches changes from two to four.  In the IC phase the wave numbers of oscillations (Fermi momenta) evolve smoothly in the range $k_{\s F,1/2} \in [\pi/2, k_{max}]$, cf. \eqref{kmax}.
At the third critical point the roots $z_\pm$ become complex conjugate:
\begin{equation}
\label{Zpm3}
  \mu = \mu_{c,2}:~ z_\mp=e^{\pm 2 ik_{max}}~.
\end{equation}
(4). In the second gapped phase (plateau $\bar{n}=1$) four roots \eqref{Z0zpm} smoothly move away from the unit circle, but stay complex, such that
\begin{equation}
\label{Zpm4}
  \mu > \mu_{c,2}:~ 1/z_\pm= z_\mp^\ast ~.
\end{equation}
This gapped phase is oscillating, with the IC wave number of oscillations decaying smoothly from the value $k_{max}$ at the phase boundary $\mu = \mu_{c,2}$,
see Fig.~\ref{KappaQ}.
This phase is similar to the oscillating gapped phase of the transverse $XY$ chain \cite{McCoyII:1971}.

We denote the real and complex parts of $k \in \mathbb{C}$ as
\begin{equation}
\label{qkap}
  k \equiv q + i \kappa,
\end{equation}
and define two critical indices\footnote{\label{Nupr} The critical index $\nu^\prime$ in different notation was first introduces in \cite{Nussinov:2012}}
near the points which are complex zeros of the partition function at $T \to 0$  as
\begin{eqnarray}
  \label{qc}
  q- \frac{m}{n} \pi \equiv \delta q  &\propto& |g-g_c^\prime|^{\nu^\prime} \\
  \kappa &\propto& |g-g_c|^{\nu}~.
  \label{kapc}
\end{eqnarray}
Here $g$ is some parameter controlling the distance to the critical point. The points $g_c^\prime$ where
the IC oscillations with the wave number $\delta q$ set in, are called disorder lines \cite{Stephenson-I:1970,*Stephenson-II:1970,*Stephenson:1970PRB}.
$g_c$ is the quantum critical point where the correlation length $\xi \equiv 1/\kappa$ diverges. In general $g_c \neq g_c^\prime$ and
$\nu \neq \nu^\prime$. Taking the modulated transverse quantum XY chain as an example, one can check the previous statement by inspection of the model's
ground-state phase diagram and the critical indices. This is also the case for the ladder at $\mu =\mu_{c,2}$, cf. Eq.~\eqref{qC2}.
The disorder lines and the lines of quantum phase transitions could merge only at some special points \cite{Chitov:2022DL}.
The representative results for $q$ and $\kappa$ are shown in Fig.~\ref{KappaQ}.

%
%
%
\section{Generalizations: Scaling and order parameter for a metallic (gapless) phase in d-dimensions}\label{Gen}
%
%
%
%
\subsection{Free fermions}\label{GenFree}
The lessons we learned from the toy ladder model allow us to make some important generalizations for the gapless (metallic) phases of $d$-dimensional
fermions. All the relevant formulas and derivations for the non-relativistic and relativistic (Dirac) fermions are presented in Sections \ref{AppNR} and \ref{AppDir} of  Appendix \ref{AppDFermi}, respectively.

For fermions with different spectra considered, a phase transition can occur only at $T = 0$. It is signalled by zeros of the partition function in the
range of real values $\mathbf{k} \in \mathbb{R}^d$ in the reciprocal (momentum) space at finite volume $V$, cf. notations \eqref{qkap} where $k \equiv |\mathbf{k}|$. Those zeros coincide with manifold of zeros of the inverse Matsubara single-particle Green's function. At $T=0$ the latter
defines the $(d-1)$-dimensional FS which separates the occupied and empty states. It is the surface of quantum criticality, and its appearance or restructuring constitutes a Lifshitz quantum phase transition.

We propose to introduce the order parameter $\mathcal{P}$ for the gapless metallic phase as a $d$-volume of the Fermi sea in the momentum space. The goal of the following analysis is to demonstrate that such bold proposal is consistent with the scaling  properties of other physical parameters, and in particular, with all scaling and hyperscaling relations, known from
the theory of quantum phase transitions, see, e.g., \cite{Imada:1998,Sachdev:2011}.

\textit{\underline{Chain, ladder:}}  Whether we consider the chain limit ($t_\perp=0$) of the Hamiltonian \eqref{Hmf}, or the ladder with three critical points,
we infer from the results of Sections \ref{Res} and \ref{LYZsec} that the order parameter near the critical points ($\mu_{c,1},\mu_{c,2}, \mu_L$)
on the gapless side behaves as:
\begin{equation}
\label{KfBeta}
  \mathcal{P} = \delta k_{\s F} \propto \sqrt{|\mu -\mu_c|}=|\mu -\mu_c|^\beta, ~~\mathrm{where}~~  \delta k_{\s F} \equiv |k_{\s F}-\frac{\pi}{2}|
\end{equation}
which, together with the results (\ref{chi12},\ref{chiL}) for compressibility,  determines two critical indices:
\begin{equation}
\label{betagamma}
  \beta = \gamma =1/2
\end{equation}
Since $\mathcal{P}$ is defined as the measure of the full Fermi sea, it adds up  ($\mathcal{P}=\mathcal{P}_1 +\mathcal{P}_2$) the contributions of two disconnected Fermi seas when $\mu_L<\mu<\mu_{c,2}$. From the scaling relation
\begin{equation}
\label{FisherSc}
  \alpha+2\beta+\gamma =2
\end{equation}
we find
\begin{equation}
\label{alpha1}
  \alpha=1/2~.
\end{equation}
At zero temperature the critical index of the heat capacity $\alpha$ is usually identified from the scaling of the ground state
thermodynamic potential $\Omega=-PV$:
\begin{equation}
\label{Palpha}
  P \propto |\mu-\mu_c|^{2-\alpha}~,
\end{equation}
see, e.g. \cite{Imada:1998,Fradkin:2013}, whence one can recover \eqref{alpha1}.
It is instructive however to make an independent check of \eqref{alpha1}

The behavior of specific heat, i.e., the derivative of the entropy near the thermal phase transition
has its counterpart for the quantum transition at $T=0$, and it is related to the entanglement.
For reviews on entanglement and various measures to quantify it, see, e.g.,  \cite{Amico:2008,Laflorencie:2016,Nishioka:2018}.
We find it most convenient to use the global entanglement \cite{WeiGold:2005,WeiGold:2011}.
It measures proximity of the given quantum state to a probe factorized (disentangled or ``classical") state.
Taking the global entanglement $\mathcal{S}$ as a counterpart of the thermodynamic entropy, we introduce the entanglement capacity
\begin{equation}
\label{EnCap}
  \mathfrak{C} \equiv \mu \frac{\partial \mathcal{S} }{\partial \mu}~,
\end{equation}
which then should have the critical properties of the specific heat. For the $XX$ chain in transverse field which belongs to the same universality class as the model \eqref{Hmf}, the global entanglement was calculated analytically in \cite{WeiGold:2005,WeiGold:2011}, leading to
\begin{equation}
\label{Cxy}
  \mathfrak{C}   \propto \frac{1}{\sqrt{|\mu-\mu_c|} }~,
\end{equation}
in agreement with \eqref{alpha1}.
The exact formula for the von Neumann entropy for a block of $L$ sites in the XX chain with transverse field \cite{Korepin:2004}
\begin{equation}
\label{Kor2004}
  \mathcal{S}=\frac{\mathcal{L}}{\pi} \ln\frac{\pi}{\mathcal{L}}~, ~~\mathcal{L} \equiv 2L \sqrt{1-h^2}=2L \sin k_{\s F}~,
\end{equation}
valid for $\mathcal{L}<1$ ($L<\xi_I$), can be also tried to calculate the entanglement capacity \eqref{EnCap} with $\mu \leftrightarrow h$, $\mu_c \leftrightarrow h_c=1$.
The result for $\mathfrak{C}$ at $h \to 1^-$ ($k_{\s F} \to 0$) and fixed $L$ agrees with the critical scaling \eqref{Cxy} up to a logarithmic factor.\footnote{\label{limits} This is the limit $L \ll \xi_I$ when the finite $\xi_I$ is the largest scale. The same limit yields the critical index $\eta^\prime$, cf. \eqref{Deta}. Note that the correlation length $\xi$ is infinite throughout the gapless phase.}

An additional comment is in order. As an extra consistency check of the above scaling analysis of the entanglement, let us recall some key results for the solvable deformation of the transverse-field XX chain with modulations, when the spin anisotropy is added, i.e., XX $\rightarrowtail$ XY. This model's deformation breaks  $U(1)$ symmetry,
resulting in appearance of anomalous terms $\propto~cc +h.c.$ in its fermionic representation, i.e., number of particles is not conserved anymore. As a consequence the IC gapless (metallic) phase vanishes and the plateaux get smeared. From the results for the order parameters in different phases of the XY chain (parity string order parameter, magnetization), gaps, and correlation lengthes, the 2D Ising universality class of the transitions is established \cite{Chitov:2019}.  In particular, $\alpha=0$, the specific heat of the Ising model diverges logarithmically. For the ladder this implies
\begin{equation}
\label{dEh}
  \mathfrak{C} (\epsilon) \propto - \ln |\epsilon|,~ |\epsilon| \ll 1~,
\end{equation}
where $\epsilon$ stands for the parametric distance to the quantum critical point. This again agrees with the results of the direct calculations of $\mathcal{S}$
for the transverse XY chain \cite{WeiGold:2005,WeiGold:2011}.

\textit{\underline{$d$-dimensional fermions:}} The order parameter is identified  with the volume of the $d$-dimensional Fermi sea
\begin{equation}
\label{OPFF}
  \mathcal{P}= \Omega_d k_{\s F}^d~\propto (\mu-\mu_c)^{\beta}~,
\end{equation}
and $k_{\s F}~\propto (\mu-\mu_c)^{1/2}=(\mu-\mu_c)^{\nu^\prime}$.
All the derivations and explicit formulas for the non-relativistic and relativistic (Dirac) fermions with the spectra $\varepsilon(\mathbf{k})=k^2/2m$, ($\mu_c=0$)
and $\varepsilon(\mathbf{\mathbf{k}})=\pm \sqrt{\mathbf{k}^2+m^2}$, ($\mu_c=m$), respectively, are collected in Appendix \ref{AppDFermi}.
The density of the ground-state grand canonical potential (i.e., pressure) scales as
\begin{equation}
\label{OmGen}
  P(\mu)  \propto (\mu-\mu_c)^{1+d/2} ~,
\end{equation}
and its second derivative
\begin{equation}
\label{ScSus}
  \chi_\circ \propto \mathcal{ N }(\mu) \propto (\mu-\mu_c)^{d/2-1}=(\mu-\mu_c)^{-\gamma}~.
\end{equation}
Here $\mathcal{ N }(\mu)$ is the density of states per spin or flavor at the Fermi energy $\mu = \varepsilon_{\s F}$ at $T=0$.
We found the following set of critical indices:\footnote{\label{zNote} The dynamical critical index $z$ should not be confused with the notations for the Lee-Yang zeros.}
\begin{equation}
\label{IndAll}
  \nu^\prime = \nu= \frac12~, ~~z=2~, ~~\beta =d  \nu^\prime~,~~\gamma=\alpha= 1-d  \nu^\prime~, ~~\eta^\prime =d~.
\end{equation}
They satisfy the Fisher identity \eqref{FisherSc} and the (hyper)scaling relations
\begin{equation}
\label{Hyper}
   \gamma = (2-\eta^\prime)\nu~,~~2-\alpha =(d+z) \nu~,~~~2 \beta = (d+z-2+\eta^\prime) \nu~.
\end{equation}
With the details of the derivations available in Appendix \ref{AppDFermi}, let us make few remarks about some subtle points of the scaling analysis
and the critical indices. The values of $\nu^\prime,\nu$ we get directly from behavior of the Lee-Yang zeros (cf. Eqs.~(\ref{qkap},\ref{qc},\ref{kapc})).
The index $z$ is obtained from scaling of the characteristic energy/gap near the critical point:
\begin{eqnarray}
  \label{xiInupr}
  \varepsilon_{\s F}-\varepsilon(0) &\propto&  \xi_I^{-z}~,~~ \mu> \mu_c \\
  \Delta &\propto&  \xi^{-z}~,~~ \mu< \mu_c~,
  \label{xinu}
\end{eqnarray}
where $\xi_I \equiv 1/k_{\s F}$ is the wavelength of oscillations and $\xi =1/\kappa$ is the correlation length (not to be confused).
Two indices $\beta$ and $\gamma$ follow from \eqref{OPFF} and \eqref{ScSus}. At this point the rest of indices is fixed by the scaling relations.
For instance, $\alpha$ can be obtained from the Fisher identity \eqref{FisherSc}. Note also that the scaling ansatz \eqref{Palpha}
implying $\partial/\partial T~\mapsto~\partial/\partial \mu$ at $T=0$ yields $\alpha=\gamma$ as an identity. An independent confirmation comes from
the scaling of the entanglement capacity $\mathfrak{C} \propto |h-h_c|^{d \nu-1}$, proposed in \cite{WeiGold:2011} for the $d$-dimensional case.
It agrees with our result \eqref{IndAll} (in the analysis of \cite{WeiGold:2011} $\nu =\nu^\prime$).\footnote{\label{EntFLT} Getting the asymptote for the entanglement entropy of the free $d$-dimensional fermions in the limit $\xi_I \gg L$ and $T \to 0$
(compare to Eq.~\eqref{Kor2004}) \cite{Gioev:2006} needs further work.}
To get the critical index $\eta^\prime=d$ without explicit use of the hyperscaling relations \eqref{Hyper}, we analyzed the density response function and the
density-density correlation function $\mathcal{D}(r)$. The latter probes the fluctuations of the order parameter, since $\mathcal{P} \propto \bar n$.
Near the critical point when $r \ll \xi_I$: $\mathcal{D}(r) \sim k_{\s F}^{2d}=1/\xi_I^{(d-2+z+\eta^\prime)}$, in agreement with \eqref{IndAll}.

Now, when the consistency of the critical behavior of the proposed order parameter with the scaling properties of other physical parameters is established, we
can make conclusions about the kind of the quantum transition occurring when the FS appears or changes its connectivity.
It depends on the dimension:

The scaling (\ref{OmGen},\ref{ScSus}) for $d=3$ agrees with the original result due to Lifshitz \cite{Lifshitz:1960}, reported for a generic spectrum (i.e., for a generic FS). The response $\chi_\circ \to 0$ at $\mu \to \mu_c \pm 0$,  and the transition is of the third order with the divergent
$P^{\prime \prime \prime} (\mu)  \propto (\mu-\mu_c)^{-1/2}$ according to the Ehrenfest classification  \cite{Lifshitz:1960}.

In two dimensions $\gamma=\alpha=0$. The density response undergoes a finite discontinuity between the incompressible gapped and
the metallic phases,  $\chi_\circ \propto \Theta(\mu-\mu_c)$, implying again the third order transition with
$P^{\prime \prime \prime} (\mu)  \propto \delta(\mu-\mu_c)$.

In one dimension the transition is of the second kind with divergent susceptibilities $ \chi_\circ \propto \mathfrak{C} \propto (\mu-\mu_c)^{-1/2}$.
The ladder model studied in this work and the modulated XX chain in transverse field are in the same universality class as the $1d$ Fermi gas.

\subsection{Interactions: Tomonaga-Luttinger and Fermi liquids }\label{GenTLFLT}
So far we did not discuss the important role of interactions in the proposed framework.
For interacting models the partition functions are unavailable in general, with the exception of handful solvable cases.
Although two paradigmatic examples of interacting fermionic systems (Tomonaga-Luttinger and Fermi liquids)  can be straightforwardly incorporated in
our analysis.

In $1d$ an arbitrarily weak repulsion is known to destroy the free-gas description (Fermi liquid theory, FLT) \cite{Giamarchi:2004}.
The low-energy sector of the interacting $1d$ fermions can be mapped onto the Hamiltonian of the free massless scalar boson field, yielding the gapless Tomonaga-Luttinger liquid (TLL).  The bosonic partition function $\mathcal{Z}$ is readily available \cite{Giamarchi:2004}, and the poles of $\mathcal{Z}$ signal that the left/right Fermi momenta are the points of quantum criticality at $T=0$, the same result as for free fermions.

Due to the Luttinger theorem the volume of the $d$-dimensional Fermi sea does not change by interactions \cite{Luttinger:1960}. 
Zeros of the inverse exact Green's function  define the interacting FS at $T=0$ \cite{AGD:1963}.  The FS cannot be interpreted anymore as a $(d-1)$-dimensional membrane in the $d$-dimensional momentum space which separates the occupied and empty states of the interacting \textit{particles}, since their distribution
$n_{\s F}(\mathbf{k})$ is not a step-function: $n_{\s F}(\mathbf{k}) \neq 0$ at $k> k_{\s F}$, it has a discontinuity $Z<1$ at $k_{\s F}$ ($d=2,3$), which becomes just an essential singulary in $1d$, since the residue of the fermionic Green's function $Z$ vanishes \cite{AGD:1963,Giamarchi:2004}.

Recall that for the non-interacting case: $\mathcal{Z}=0 \Leftrightarrow \mathrm{det} \hat G^{-1}=0$, whence the equations for the Lee-Yang zeros \eqref{ZGrNr} and \eqref{LYZD} follow exactly. In case of interactions no such simple relation is available, and in general the equations for zeros of the partitions function and for zeros of the inverse Green's functions are distinct. 
From the rigorous results relating the interacting thermodynamic potential ($\Omega$) and the exact temperature Green's function \cite{LuttingerWard:1960,AGD:1963}, the average particle density is:
\begin{equation}
\label{NGr}
 -\frac{1}{V} \frac{\partial \Omega}{\partial \mu }=\bar n = T \sum_{\omega_n} \int \frac{d \mathbf{k}}{(2\pi)^d } G(\mathbf{k},\omega_n)e^{ i \omega_n 0^+}
 =\frac{\Omega_d}{(2 \pi)^d}  k_{\s F}^d    ~.
\end{equation}
The above equation along with the exact expression for the thermodynamic potential which contains (under summation and integration) the term $\propto \ln G(\mathbf{k},\omega_n)$ \cite{LuttingerWard:1960,AGD:1963}, imply that the zeros of the inverse exact Green's function still define the quantum critical manifold where $\Omega$ has a non-analyticity.

Thus, appearance of the FS is a quantum transition in the liquid of interacting fermions. The FS corresponds to the special points of the partition function $\mathcal{Z}$ where $\Omega$ has non-analyticity, this $(d-1)$-dimensional membrane preserves the volume inside it, although becomes penetrable, and it is a locus on the gapless modes.

The renormalization-group (RG) analysis of the low-energy effective action of interacting fermions in $d=2,3$ with isotropic spectrum at $T \neq 0$, demonstrates that the Landau Fermi liquid is robust, proviso that the Pomeranchuk stability conditions are satisfied \cite{Chitov:1995}. Note that the limit $T \to 0$ in the RG equations is subtle, and the whole FLT can be missed if the RG flow is attempted to be traced directly at $T=0$ \cite{Shankar:1994}. The narrow angular anomalies in the Landau interaction function and the scattering vertex, a.k.a the non-analytic temperature corrections to the FLT, become discontinuities while approaching the quantum critical manifold at $T \to 0$  \cite{Chitov:1998}.

To obtain the scaling exponents for the Fermi liquid we first use the Luttinger theorem to relate the spectrum of the non-interacting fermions and the Fermi energy. As a result we recover the critical indices $\nu^\prime,z$ of the free-fermionic case \eqref{IndAll}. The
index $\beta= d \nu^\prime$ follows directly from geometry. The compressibility \eqref{chi0expl} of the Fermi liquid is renormalized by interactions:
\begin{equation}
\label{chi0FLT}
  \chi_0= \frac{\partial \bar{n} }{\partial \mu} =\frac{d \Omega_d}{(2 \pi)^d} m \frac{1+F_1}{1+F_0} k_{\s F}^{d-2}~,~~d=2,3,
\end{equation}
where $F_{0,1}$ are the components of the expansion of the angular-dependent Landau interaction function $f$ into the series over Legendre polynomials ($d=3$)
or Fourier harmonics ($d=2$), see, e.g., \cite{AGD:1963,Chitov:1995}.
The interactions generate additional corrections to the perfect scaling law $\chi_0 \propto k_{\s F}^{d-2}$ in $3d$, since the dimensionless couplings  $F_{0,1}$ in \eqref{chi0FLT} are defined via the dimensionful Landau interaction  function $f$ multiplied by the density of states at the Fermi level $\mathcal{N}(\mu) \propto k_{\s F}^{d-2}$. Those corrections are irrelevant near the critical point  $k_{\s F} \to 0$, yielding the non-interacting index $\gamma$. With the indices $\nu^\prime,z,\beta,\gamma$, the rest follows from the (hyper)scaling relations, and we end up with the result \eqref{IndAll} for $d=2,3$.

Before we proceed, one refinement not related to the interactions \textit{per se}, needs to be done. In the standard derivations of the FLT or in the above discussions, it is assumed that we are dealing with one fermionic band, which is partially filled with the Fermi sea. In case of lattice fermions with several bands (like, e.g.  the ladder considered in Sec.~\ref{LadMod}), the Luttinger theorem \eqref{NGr} needs to be stated in a more general way to account for the contribution into the total particle number $\bar N$ from the fully filled bands $\bar n_i$:
\begin{equation}
\label{LuttGen}
 \bar n =\frac{\bar N}{V}= \sum_{i} \bar n_i +\frac{\Omega_d}{(2 \pi)^d}  k_{\s F}^d    ~,~T=0.
\end{equation}
In case of disconnected Fermi seas, their $d$-volumes must be added on the r.h.s. of \eqref{LuttGen}. In general those contributions could be more complicated
than the volume of $d$-ball explicitly used in \eqref{LuttGen}. The relevant point is that turning on interactions does not change the total volume of the Fermi sea.

In $1d$ interactions destroy the Fermi liquid, but the Fermi sea survives and even preserves its volume in the $k$-space. To follow up the role of interactions
let us consider the antiferromagnetic XXZ Heisenberg chain in the transverse field within the interaction (anisotropy) range $-1 \leq \Delta \leq 0$, $\Delta \equiv -J_z/J$.
The model is integrable, and its phase diagram is known \cite{Takahashi:1999}. In the free fermionic limit $\Delta=0$ it is the transverse XX chain. The latter is the limit $\delta=t_\perp=\mu_a=0$ of the model \eqref{Hmf}. Interactions in the range $|\Delta| \leq 1$ do not change qualitatively the phase diagram of the XXZ chain in transverse field with respect to its free fermionic XX counterpart. It is in agreement with the plateau theorem  \eqref{plateaueq}.
Without modulations in the chain ($P=1$) the first gapped phase (plateau at $\mathcal{M}=0$) is absent, while the critical field $h_c=1-\Delta$ for the plateau at $\mathcal{M}=1/2$ in the polarized gapped phase, is renormalized due to interactions. Compare to Fig.~\ref{BandsPl}c. However, the fundamental difference is
that the gapless IC phase of the XXZ chain in the range $0 \leq h \leq h_c$ is the TLL. Magnetization near the Lifshitz critical point $h_c$ where the Fermi sea vanishes ($k_{\s F} \to 0$), is known exactly \cite{Korepin:1986}:
\begin{equation}
\label{MzXXZ}
  \mathcal{M} \approx \frac12 - \frac{\sqrt{2}}{\pi} \sqrt{h_c-h}~,~~h \to h_c^-~.
\end{equation}
Up to the critical field renormalization, it is the same result as for free fermions. It yields $\nu^\prime =1/2$ and the divergent susceptibility with
$\gamma =1/2$. At low filling  $\varepsilon_{\s F} \propto  k_{\s F}^2$, which gives $z=2$. So the Lifshitz transition $2 \to 0$ in the TLL is
in the universality class of the free fermions with the critical indices \eqref{IndAll} for $d=1$.

The bosonization of the Tomonaga-Luttinger model of spinless fermions yields the density-density correlation function \cite{Giamarchi:2004}:
\begin{equation}
\label{DDBos}
  \langle \rho(x)\rho(0) \rangle \propto \frac{1}{x^2}+~\mathrm{oscillating ~terms}~,
\end{equation}
which can be used as an additional cross-check. In contrast to its counterpart for free fermions, it contains a perfectly scale-free first term. By matching  $2d=d+z-2+\eta$ we recover $\eta=1$.

The value $\alpha=1/2$ of the TLL obtained from the scaling relations warrants a special comment. The conformal field theory $(d=1+1)$ result for the entanglement entropy (see \cite{Nishioka:2018} for different derivations and original references)  is:
\begin{equation}
\label{SCFT}
  \mathcal{S}(L)=\frac{c}{3} \ln \frac{L}{\epsilon}+~...~,
\end{equation}
where $c=1$ is the central charge of the free bosonic field and $\epsilon$ is the UV cutoff. With $\epsilon \sim \xi_I$ it agrees with the exact result of Jin and Korepin \cite{Korepin:2004} for the XX chain in the limit $L \gg \xi_I$. According to the results of Vidal \textit{et al} \cite{Vidal:2003}, the XX (free fermions) and XXX (TLL) chains demonstrate the same universal behavior \eqref{SCFT} with $c=1$ far from the Lifshitz point in the gapless phase, when one can probe the regime $L \gg \xi_I$.

The Lifshitz point of the transverse XX chain at $h_c$ is a special point, where the lines of quantum criticality and the disorder line ($\xi_I \to \infty$) merge. The latter is also the disentanglement line. (At merging critical lines the limiting value of the entanglement entropy depends on the direction to approach the critical point and on the order limits are taken \cite{Franchini:2007,Franchini:2011}.)
The entanglement measured by different quantities: concurrence \cite{Chitov:2022DL}, global entanglement \cite{WeiGold:2005,WeiGold:2011}, and von Neumann entropy \cite{Korepin:2004} (cf. \eqref{Kor2004}), vanish at $h_c$.\footnote{\label{DLEnt} The Lifshitz critical point $\mathcal{P} \to 0$ is not necessarily the disorder point where $\xi_I \to \infty$. The example is the two-leg ladder at the critical point $\mu_{c,2}$, cf. Eq.~\eqref{qC2}. The Fermi sea vanishes not at the center/edge of the BZ ($\Im z_\pm =0$ in such case), but somewhere at an IC point $k_{max}$ where $\Im z_\pm(h_{c,2}) \neq 0$. See Figs.~\ref{ZRoots} and \ref{KappaQ}.
The IC oscillations continue into the gapped phase. The scaling relations are preserved, while from the results of \cite{Chitov:2022DL} we expect a residual entanglement at $\mu_{c,2}$ which does not affect the divergence of $\mathcal{C}$ as in Eq.~\eqref{Cxy}.}
The exact results on the factorizing field of the transverse XXZ chain \cite{Muller:1982} confirm that the critical field $h_c$ is also the field of disentanglement (disorder line). Then the scaling relations and the available non-interacting result \eqref{Kor2004} \cite{Korepin:2004} imply that the entanglement of the XXZ chain ($L<\xi_I$) vanishes at $h_c=1-\Delta$ as
\begin{equation}
\label{EntXXZ}
 \mathcal{S} \propto - (h_c-h)^{1/2} \ln(h_c-h) ~,~~h \to h_c^-~.
\end{equation}
More generally, we conjecture that if the Lifshitz point of the gapless $d$-dimensional fermionic liquid is also the disorder point where $\xi_I \to \infty$, then the entanglement $\mathcal{S}$ vanishes from the gapless side as
\begin{equation}
\label{Sgen}
  \mathcal{S} \propto |\mu -\mu_c|^{1-\alpha}~,
\end{equation}
(possibly up to a logarithmic factor) with $\alpha$ given by \eqref{IndAll}.

%
%
%
\section{Conclusion and discussion}\label{Concl}
%
%
%

In this work we studied critical properties and ground-state phases of several models of $d$-dimensional fermions.
We used the Yang and Lee approach to detect a phase transition through the search of zeros of the model's partition function.
For the fermions with different spectra considered, a phase transition can occur only at $T = 0$. At the critical point zeros of the partition function are in the
range of real values $\mathbf{k} \in \mathbb{R}^d$ in the momentum space at finite volume $V$.
The manifold of those zeros at $T=0$ defines the $(d-1)$-dimensional FS. It defines the surface of quantum criticality, and its appearance or restructuring constitutes a quantum (Lifshitz) phase transition.
Appearance of such a membrane in the momentum space \cite{Horava:2005} breaks its initial symmetry, and as a consequence gapless excitations around the FS appear as well, which is a hallmark of metallic phase.

As an exactly solvable simplified version of the dimerized spin-$\frac12$ Heisenberg ladder in transverse magnetic field,
we studied the spinless fermions on a two-leg ladder. This tight-binding one-dimensional model has the ground state properties consistent with the general plateau theorem and, an addition, it undergoes a gapless-to-gapless Lifshitz transition within the IC (metallic) phase. It is an interesting model to study exactly the metal-to-metal
and the metal-insulator transitions. 
Analysis of the Lee-Yang zeros in this $1d$ model is conveniently done when the wavenumbers from the BZ are mapped onto a unit circle in the complex plane.
From that analysis we get the information about gapped and gapless phases, by tracing positions of zeros with respect to the unit circle. In particular,
the number of the Fermi points and its doubling at the Lifshitz transitions is read off from the number of the roots sitting on the unit circle.
The analytical expressions for the Lee-Yang zeros yield directly various thermodynamic parameters (e.g., characteristic lengths, particle density, susceptibilities).

The result derived for the ladder (cf. Eq.~\eqref{nkpm} in the Appendix, reproduced below for convenience)
\begin{equation}
\nonumber
  \bar{n}= 1 - \frac{1}{4 \pi} ( \arg z_+ - \arg z_-)~,
\end{equation}
relating the average occupation number to the Lee-Yang zeros $z_\pm$ is suggestive: it implies that for a gapless (metallic) phase we can do a little better
than just count the number of the Fermi points (Lee-Yang zeros) on the different sides of the Lifshitz transition, but rather introduce a continuous (order) parameter which quantifies this phase in terms of the ``depth'' of its Fermi sea.
Following this logic, the order parameter $\mathcal{P}$ for the gapless metallic phase relating it to the $d$-volume of the Fermi sea in the momentum space
is introduced for the first time.

To cross-check consistency of this proposal we analysed the free $d$-dimensional fermions for $d=1,2,3$ with the Dirac and non-relativistic spectra.
We demonstrate that for all models considered, the properties of $\mathcal{P}$ are consistent with the scaling of other physical parameters, like correlation length, the IC (Friedel) oscillation wavelength, compressibility, and the entanglement. Taking the global entanglement (or entanglement entropy) $\mathcal{S}$ as a zero-temperature counterpart of the thermodynamic entropy, we introduced in this work the entanglement capacity which should have the critical properties of the specific heat with the critical index $\alpha$. It is shown that the value of $\alpha$ obtained from our scaling analysis agrees with the available results for entanglement.
Also, all the scaling and hyperscaling relations for critical indices, known from the theory of quantum phase transitions, are shown to be satisfied. The Lifshitz transition related the appearance or restructuring of the FS is found to be of the third kind in three and two spatial dimensions, and it is of the second kind with divergent susceptibilities in one dimension. The scaling indices and the third order of transition we found for $d=3$ agree with the original result due to Lifshitz \cite{Lifshitz:1960}, reported for a generic spectrum (i.e., for a generic FS).

To address the important role of interactions, two paradigmatic examples of interacting \textit{gapless} fermionic systems (Tomonaga-Luttinger and Fermi liquids) are analyzed. Using the Luttinger theorem along with other known results obtained by the methods of quantum field theory, RG, bosonization, it is shown that the
stable Landau Fermi liquid belongs to the same universality class (defined by the critical indices) as the free Fermi gas in $d=2,3$.
In $1d$ the repulsive interactions drive the fermions to the TLL state which is shown to possess the universality class of the $1d$ Fermi gas. It is quite remarkable that the scaling properties of the TLL parameters near the point $ k_{\s F} \to 0$ of transition related to the FS order parameter $\mathcal{P}$ is universal, while many other quantities and correlation functions are known from the exact results to be non-universal: exponents of the power-law functions explicitly depend on couplings. We think that such universality stems from invariance of the Fermi sea volume with respect to interactions (the Luttinger theorem) across all dimensions $d=1,2,3$, and from the geometric (or topological) nature of $\mathcal{P}$.

The FS is the manifold where the exact inverse Matsubara single-particle fermionic Green's function $G^{-1}$ vanishes in the limit $T \to 0$. The available rigorous results for the thermodynamic potential $\Omega$ and the present analysis corroborate the statement that the FS is the quantum critical manifold where $\Omega$ has non-analyticities, but the rigorous proof relating special (singular) points of the partition function $\mathcal{Z}$ to the exact Green's function for a general interacting case is still lacking. The $d=1$ case is an exception: the Fermi points can be identified by the poles of the partition function of the Tomonaga-Luttinger model in its representation of the free bosonic field.

Although the volume of the Fermi sea (i.e., the order parameter $\mathcal{P}$ for the gapless phase) is unaltered by interactions, the role of interactions in the stability of the gapless phase, its boundaries and other properties near transition points, need to be further studied case by case. This is left beyond the scope of the present work.

An interesting correspondence between ideal Bose and Fermi gases  can be established. One can easily check from a textbook analysis of the paradigmatic Bose-Einstein condensation (BEC) \cite{LandauV5} that this phase transition accompanied by appearance of the condensate, is signalled by
a \emph{pole} of the partition function, when the chemical potential of the gas vanishes at the critical temperature of BEC. The ideal Fermi gas undergoes a phase transition at $T=0$: the Fermi sphere appears as a manifold of \emph{zeros} of the partition function.
\begin{equation}
\label{BECFS}
\nonumber
 \mathcal{Z} \longrightarrow
  \left\{
  \begin{array}{lr}
  \infty~~~ &\mathrm{BEC~~(Bose)}  \\ [0.2cm]
  0~~~      &\mathrm{FS~~~(Fermi)}\\
  \end{array}
  \right.
\end{equation}
This brings a certain harmony between two types of ideal quantum gases, alleviating misconception that only the Bose gas can undergo a phase transition.

%
\begin{acknowledgments}
I thank Z. Nussinov for a helpful discussion and for bringing important references to my attention.
The author acknowledges financial support from the Russian Science Foundation (RSF),
grant No.~24-22-00075, https://rscf.ru/24-22-00075.

\end{acknowledgments}

\begin{appendix}
\section{Lifshitz transitions in two-leg ladder model}\label{AppLad}
%
%
We start from the Heisenberg spin-$\frac12$ two-leg ladder:
\begin{equation}
\label{Ham}
  H_{S}=\sum_{\alpha=1}^{2}\sum_{n=1}^{N}\Big\{J_{\alpha}(n)
  \mathbf{S}_{\alpha}(n)\cdot\mathbf{S}_{\alpha}(n+1)
  +h_{\alpha}(n)S_{\alpha}^{z}(n)  \Big\}
  +\sum_{n=1}^{N} J_{\perp}(n)
  \mathbf{S}_{\alpha}(n)\cdot\mathbf{S}_{\alpha+1}(n).
\end{equation}
The spin operators $\mathbf{S}= \frac12 \boldsymbol{\sigma}$ are defined via the standard Pauli matrices $\boldsymbol{\sigma}$,
all spin exchange couplings are assumed antiferromagnetic.  The ladder is intrinsically dimerized along the chains with alternating coupling
$J_{\alpha}(n)=J[1+(-1)^{n+\alpha} \delta_S]$, see Fig.~\ref{SDLadder}. The dimerization parameter lies within the range $\delta_\sharp \in [0,1]$.
The transverse magnetic field has uniform and staggered components $h_{\alpha}(n)= h+(-1)^{n+\alpha}h_a$.

When the spin ladder \eqref{Ham} is fermionized and the exact interacting Hamiltonian of spinless fermions is treated within the Hartree-Fock approximation,
the problem reduces to dealing with the quadratic effective fermionic Hamiltonian \eqref{Hmf}.

In the mean-field approach, couplings of the fermionic Hamiltonian \eqref{Hmf} are renormalized due to interactions. They are some involved functions of the microscopic parameters of the spin model \eqref{Ham}, to be determined self-consistently.\footnote{\label{NoteMFA} We infer from earlier work \cite{Azzouz:1993,*Azzouz:1994,Dai:1998,Hori:2004,Nunner:2004,Chitov:2007,Chitov:2008,Chitov:2017JSM,Chitov:2020,Toplal:2022,Chitov:2022} where technical details can be found, that such mean-field approach is quite efficient, even quantitatively, for analysis of quantum chains and ladders. In particular, it reproduces the plateaux structure in accord with the theorem \eqref{plateaueq}, and the predicted quantum critical points \cite{Toplal:2022} are in decent numerical agreement with the available  Monte Carlo and exact diagonalization results.}
Roughly, $t \propto J$, $\delta \propto \delta_S$, $\mu \propto h$, etc. Their exact values are not important for the following, it is just a technicality. What matters is that the ground-state phases and quantum transition between them, i.e., the physics we want to explore is fully imbedded in the Hamiltonian \eqref{Hmf}, which can be viewed as a ``light" solvable version of \eqref{Ham}.

\subsection{Plateaux and gapless-to-gapless Lifshitz transition in the IC phase}\label{AppL}
The spectrum  \eqref{spectra_ss1} reveals two gapped phases with plateaux connected to a gapless IC phase.
In the limiting case $t_\perp =0$ or $\delta =0$, the bands merge: $\varepsilon_+(k)=\varepsilon_-(k) \equiv \varepsilon(k)$. In this case two plateaux are connected by a \textit{smooth} continuous curve of $\bar{n} \in \mathbb{R}$ in the IC gapless phase, as shown in Fig.~\ref{BandsPl}c.

When two dimerized ($\delta \neq 0$) chains are coupled in a ladder ($t_\perp \neq 0$),
the band structure reveals a Lifshitz transition within the gapless phase.
Two bands $\varepsilon_\pm(k)$  are split, even if the staggered potential $\mu_a$ is absent, see Fig.~\ref{BandsPl}b.
The system stays in the gapped phase (plateau $\bar{n}=\frac12$) at $0 \leq \mu < \mu_{c,1}$.  At the critical value
\begin{equation}
\label{hc1}
 \mu_{c,1} =\sqrt{\mu_a^2+\big(\delta-t_\perp \big)^2}~,
\end{equation}
the gap closes at the edges of the BZ and a continuous transition into the gapless IC phase takes place.

The band is partially filled, and the filled states are localized between the edges of the BZ $k=\pm \pi/2$ and two Fermi points $\pm k_{\s F, 1}$ defined by the equation
\begin{equation}
\label{kF1}
  \mu  =\varepsilon_\mp (\pm k_{\s F,1})
\end{equation}
When growing $\mu$ reaches the critical value $\mu_{ L}$, another couple of gapless points in the spectrum appears at the edges of the BZ:
\begin{equation}
\label{hL}
 k_{\s F,1}= k_{L}:~~ \mu_{ L}=\varepsilon_\pm (k_{L})=\varepsilon_\mp  \big(\mp \pi/2 \big)= \sqrt{\mu_a^2+\big(\delta + t_\perp \big)^2}~,
\end{equation}
with
\begin{equation}
\label{kL}
  k_L= \pm \Bigg(  \frac{\pi}{2} +2 \arcsin \sqrt{\frac{\delta t_\perp}{1-\delta^2}} \Bigg)~.
\end{equation}
This is another Lifshitz transition: number of the Fermi points doubles $\{\pm k_{\s F,1} \} \rightarrow \{ \pm k_{\s F, 1}, \pm k_{\s F,2} \}$, yielding a cusp of  $\bar{n}(\mu)$ in the IC phase, seen in Fig.~\ref{BandsPl}d.
When $\mu$ reaches the upper critical value
\begin{equation}
\label{hc2def}
 \mu_{c,2} = \max \varepsilon_\pm ~,
\end{equation}
where the maxima of $\varepsilon_\pm$ occur at  $k_{max}$, both bands become totally filled,
and a quantum transition of the second order occurs into the fully filled $\bar{n}=1$ (polarized)
gapped phase. Explicitly:
\begin{equation}
\label{kmax}
  k_{max}=  \pm \arcsin \frac{\delta t_\perp}{1-\delta^2}~,
\end{equation}
and
\begin{equation}
\label{hc2}
 \mu_{c,2} =\sqrt{1+\mu_a^2+ \frac{t_\perp^2}{1-\delta^2} }~.
\end{equation}
The gaps in the spectrum of the Hamiltonian close/open linearly near $k =\pm \pi/2$ or $\pm k_{max}$ at two critical points at the edges of plateaux:
\begin{equation}
\label{Gap}
  \Delta =|\mu- \mu_{c,\sharp}|~~ \longrightarrow~~ z \nu =1~.
\end{equation}
At the critical points  $\mu_{c,1/2}$ the excitation spectrum is parabolic:
\begin{equation}
\label{Parab}
  \varepsilon_\pm(k) -\mu_{c,1/2} \propto (k-k_\sharp)^2~,
\end{equation}
where $k_\sharp =\frac{\pi}{2}$ and $k_\sharp = k_{max}$ for $\mu_{c,1}$ and $\mu_{c,2}$, respectively.
In the gapless phase at $\mu_{c,1}< \mu < \mu_L$ the excitation spectrum is linear near $k_{\s F,1}$ with the Fermi
velocity $v_{\s F,1}$, while at $\mu_L <\mu  < \mu_{c,2}$ two linear spectra have different Fermi velocities $v_{\s F,1} \neq v_{\s F,2}$.
They are defined as:
\begin{equation}
\label{vF12}
   v_{\s F,1/2}= \Big|   \frac{\partial \varepsilon_\mp(k)}{\partial k} \Big|_{k_{\s F,1/2}}
\end{equation}
The occupation number is found as:
\begin{equation}
\label{NMzAn}
 \bar{n}- \frac12 =
  \left\{
  \begin{array}{lr}
  \frac12~,~ &\mu > \mu_{c,2} \\ [0.2cm]
  \frac12  - \frac{1}{2\pi}( k_{\s F,1} + k_{\s F,2})~, ~&\mu_L <\mu  < \mu_{c,2} \\ [0.2cm]
  \frac14 - \frac{1}{2 \pi} k_{\s F,1}~, ~&\mu_{c,1}< \mu < \mu_L \\ [0.2cm]
  0~.~ &\mu < \mu_{c,1}\\
  \end{array}
  \right.
\end{equation}
The roots of $E_\alpha(k)=0$ yield two Fermi points:
\begin{equation}
\label{ypm}
  \sin k_{\s F,2} = \frac{\delta t_\perp}{1-\delta^2} \pm \sqrt{ \frac{\mu _{c,2}^2-\mu^2}{1-\delta^2}  } \equiv y_\pm~,
\end{equation}
while the other couple of points $k_{\s F,1}$ is given by the above expression with the substitution $t_\perp \rightarrow -t_\perp$,  $k_{\s F,2} \rightarrow k_{\s F,1}$,
cf. Eq.~\eqref{spectra_ss1}.
From \eqref{NMzAn} we easily find
\begin{equation}
\label{chi}
  \chi_\circ = \frac{\partial \bar{n} }{\partial \mu}=
  \left\{
  \begin{array}{lr}
  0,~ &\mu > \mu_{c,2} \\ [0.2cm]
  -\frac{1}{2\pi} \Big( \frac{\partial k_{\s F,1} }{\partial \mu} + \frac{\partial k_{\s F,2}}{\partial \mu} \Big),
  ~&\mu_L <\mu < \mu_{c,2} \\ [0.2cm]
   - \frac{1}{2 \pi} \frac{\partial k_{\s F,1} }{\partial \mu}, ~&\mu_{c,1}< \mu < \mu_L \\ [0.2cm]
  0,~ &\mu < \mu_{c,1}\\
  \end{array}
  \right.
\end{equation}
which is a direct counterpart of the spin susceptibility $\chi = \frac{\partial \mathcal{M} }{\partial h}$, cf. \eqref{MNudef}.  The density response $\chi_\circ$ is proportional to the compressibility $\mathcal{X}$ which is rigorously defined as:\cite{LandauV5}
\begin{equation}
\label{kappa}
  \mathcal{X}= \frac{1}{ \bar{n}^2} \frac{\partial \bar{n} }{\partial \mu}
\end{equation}
The derivatives in Eq.~\eqref{chi} are given by the explicit formula:
\begin{equation}
\label{derKf}
  \frac{\partial k_{\s F,1/2} }{\partial \mu} =-\frac{\mu}{1-\delta^2}\frac{1}{ \cos k_{\s F,1/2} \big( \sin k_{\s F,1/2} \pm \sin k_{max}\big)}
\end{equation}
The compressibility is plotted in Fig.~\ref{Comp}.

\subsection{Zeros of partition function and quantum transitions}\label{AppLYZL}
The condition $\mathcal{Z}(\mu,\beta, N)=0$ for zeros of the partition function of the grand canonical ensemble amounts to \cite{Chitov:2021,Chitov:2022DL}
\begin{equation}
\label{ZGr}
  \prod_{k,n,\alpha} G_\alpha^{-1}(k,\omega_n)=0
\end{equation}
for the inverse temperature Green's function:
\begin{equation}
\label{Green}
  G_\alpha^{-1}(k,\omega_n)=i \omega_n - E_\alpha(k),
\end{equation}
where  $\omega_n=(2n+1)\pi T$ is the fermionic Matsubara frequency.
To conveniently find the Lee-Yang zeros we map the reduced BZ onto a unit circle on the complex plane $z$ as:
\begin{equation}
\label{zk}
  z \equiv e^{2i k}~.
\end{equation}
For a given $z$, the ground-state zeros of (\ref{ZGr}) are the roots of
\begin{equation}
\label{Ea0}
 \prod_{\alpha=1}^4 E_\alpha(z)=0~.
\end{equation}
Using parameters $y_\pm$ defined in Eq.~\eqref{ypm}, which are now allowed to admit any complex values, Eq.~\eqref{Ea0}
is written as:
\begin{equation}
\label{Z0zpm}
 (z-z_+)(z-z_-)(z-z_+^{-1})(z-z_-^{-1})   =0~.
\end{equation}
It has four roots, two of them  are mutually reciprocal, and
\begin{equation}
\label{ZpmY}
  z_\pm = \Big(\sqrt{1-y_\pm^2} + i y_\pm \Big)^2~.
\end{equation}
The quantum critical points correspond to the physical roots \eqref{ZpmY}, when $k \in \mathbb{R}$, i.e., those which lie on a unit circle $|z_\pm|=1$.
Evolution of the positions of the roots on the complex plane with respect to the unit circle in four different phases is shown in Fig.~\ref{ZRoots}.

With the parametrization \eqref{zk}, the correlation length and the wave number of oscillations in the ladder
are determined by the couple of conjugate roots $z_\pm$ closest to the unit circle.
The analytical solution for the complex wavenumber $k_\pm$ follows from Eqs. \eqref{zk} and \eqref{ZpmY} as
\begin{equation}
\label{ksharp}
  k_\sharp= \frac{1}{ i} \ln \Big( \sqrt{1-y_\sharp^2}+i y_\sharp \Big)=
  \arcsin y_\sharp
\end{equation}
The representative results are shown in Fig.~\ref{KappaQ}.
Near the critical points
\begin{equation}
\label{kapC}
  |\kappa_\pm| \propto \sqrt{|\mu -\mu_\sharp|} ~,~~\sharp =L,c1,c2,
\end{equation}
and
\begin{equation}
\label{qC1}
  \delta q_\pm \propto \sqrt{|\mu -\mu_\sharp|} ~,~~\sharp =L,c1,
\end{equation}
which implies along with \eqref{Gap}
\begin{equation}
\label{znu2}
  \nu^\prime = \nu = \frac12, ~~z=2.
\end{equation}
The critical point $\mu_{c,2}$ is not a disorder point, since the IC oscillations continuously (albeit not smoothly)
evolve into the filled phase  at $\mu >\mu_{c,2}$, see Fig.~\ref{KappaQ}. Near $\mu_{c,2}$
\begin{equation}
\label{qC2}
  q_\pm -  \arcsin \frac{\delta t_\perp}{1-\delta^2} \propto
   \left\{
  \begin{array}{lr}
  \pm \sqrt{\mu_{c,2} -\mu},~ &\mu \to  \mu_{c,2}-0 \\ [0.2cm]
  -(\mu -\mu_{c,2}) ,~ &\mu \to  \mu_{c,2}+0 \\
  \end{array}
  \right.
\end{equation}
The average occupation number is related to the Lee-Yang zeros as:
\begin{equation}
\label{nkpm}
  \bar{n}= 1 - \frac{1}{4 \pi} ( \arg z_+ - \arg z_-) = \frac{1}{2 \pi} (q_+-q_-),
\end{equation}
whence the compressibility \eqref{kappa} is readily found with
\begin{equation}
\label{dnmuLY}
    \frac{\partial \bar{n}}{\partial \mu}= \frac{\mu}{2 \pi \sqrt{1-\delta^2}}
    \Re \Bigg\{ \frac{1}{\sqrt{\mu _{c,2}^2-\mu^2}}
    \Bigg[ \frac{1}{\sqrt{1-y_+^2}} + \frac{1}{\sqrt{1-y_-^2}} \Bigg] \Bigg\}.
\end{equation}
Once the branch cuts are chosen to yield the signs as shown in Fig.~\ref{KappaQ}, the above expressions are applicable through the whole range $\mu \geq 0$,
reproducing straightforwardly the results for $\bar{n}$ and $\chi_\circ$ of the previous section, although in a less intuitive way than the derivations based on the band filling arguments.\footnote{\label{qkF} Note that relating $q_\pm$ to the Fermi momenta $k_{\s F,1/2}$  makes sense only in the gapless IC phase.
In particular, $\mathcal{P}=0$ at $\mu>\mu_{c,2}$, while $q_\pm$ evolves continuously through $\mu_{c,2}$.}

The first expression on the rhs of \eqref{nkpm} helps to connect the results for the Lee-Yang roots to the generalized LSM theorem \cite{Oshikawa:2000}: to get an incommensurate filling at least one of the roots $z_\pm$ must leave the real axis, while it is not sufficient, as we infer from the case $\mu > \mu _{c,2}$. The arguments of the complex $z_+$ and $z_-$ cancel, and the phase is fully filled (polarized) with $\bar{n}=1$. See Fig.~\ref{ZRoots}.

%
%

\section{$d$-dimensional fermions - Lee-Yang zeros, scaling, and correlation functions}\label{AppDFermi}
%
%

The static density-density response function, defined in the standard way \cite{Girvin:2019}
\begin{equation}
\label{chiQ}
  \chi(\mathbf{q}) = \chi_\circ L(\mathbf{q})
\end{equation}
is given for the free Fermi gas by the contribution of the particle-hole loop of the single-particle Green's functions.
The dimensionless Lindhard function is normalized such that $L(0)=1$.
The homogeneous response $\chi_\circ = \chi(0)$
\begin{equation}
\label{chi0}
  \chi_\circ = \frac{\partial \bar n}{\partial \mu}= d \mathcal{ N }(\mu)~,
\end{equation}
is equal (up to a factor $d$) to the density of states per spin or flavor:
\begin{equation}
\label{DOSdef}
  \mathcal{ N }(\mu) \equiv \int \delta(\varepsilon(\mathbf{k})-\mu) \frac{d \mathbf{k}}{(2 \pi)^d}~.
\end{equation}

\subsection{Free non-relativistic fermions}\label{AppNR}

Consider the grand canonical ensemble of non-relativistic $d$-dimensional fermions with the spectrum
$\varepsilon(\mathbf{k})=k^2/2m$  ($2m \mu(T=0) =k_{\s F}^2$, $\mu_c=0$). The Lee-Yang zeros of the partition function are the zeros
of the inverse temperature Green's function \cite{Chitov:2021}:
\begin{equation}
\label{ZGrNr}
  \prod_{\mathbf{k},n} G^{-1}(\mathbf{k},\omega_n)=0~,
\end{equation}
where
\begin{equation}
\label{GrFLT}
  G^{-1}(\mathbf{k},\omega_n)=i \omega_n - \varepsilon(\mathbf{k})+\mu~.
\end{equation}
No phase transition can occur at $T \neq 0$, since $\nexists~ k \equiv |\mathbf{k}| \in  \mathbb{R}$ to solve (\ref{ZGrNr},\ref{GrFLT}).
Using notations \eqref{qkap} for the complex solution of (\ref{ZGrNr},\ref{GrFLT}) one can easily find \cite{Chitov:2021} for $T/\varepsilon_{\s F} \ll 1$ the wave vector of the IC oscillations\footnote{\label{KZet} Note that in the analysis of the Lee-Yang
zeros of the ladder model, $z$ is just a convenient parameter to map the 1D BZ onto a unit circle $|z|=1$. The actual task is
to identify the transition point(s) as the solutions for zeros of the partition function in the real range of parameters and, in particular,
for real values $\mathbf{k} \in \mathbb{R}^d$ in the reciprocal space at finite volume $V$.}
\begin{equation}
\label{qTFLT}
  q = k_{\s F} +\mathcal{O}((T/\varepsilon_{\s F})^2)
\end{equation}
and the inverse correlation length
\begin{equation}
\label{qTFLT2}
  \kappa = \pi m T/k_{\s F}~.
\end{equation}
In the limit $T \to 0$ the roots (\ref{qTFLT},\ref{qTFLT2}) yield the manifold of the QPT points in the range of real $\mathbf{k}$:
\begin{equation}
\label{FSnr}
  \mu=\varepsilon(\mathbf{k})~,
\end{equation}
which is just the equation for the FS with $\varepsilon_{\s F}=\mu(T=0)$. For simplicity we will give the results per one
fermionic flavor (spin). Identifying the order parameter $\mathcal{P}$ with the volume of the Fermi sea as
\begin{equation}
\label{OPnr}
  \mathcal{P}= \Omega_d k_{\s F}^d~\propto \mu^{\beta}~,
\end{equation}
we have $\mathcal{P} \propto \bar{n}$, where
\begin{equation}
\label{avn}
  \bar{n}= \frac{\Omega_d}{(2 \pi)^d} k_{\s F}^d~
\end{equation}
is the average particle density and $\Omega_d$ is the volume of $d$-dimensional unit ball.
The response is
\begin{equation}
\label{chi0expl}
 \chi_\circ = \frac{\partial \bar{n} }{\partial \mu} =\frac{d \Omega_d}{(2 \pi)^d} m k_{\s F}^{d-2}~.
\end{equation}
Using the above results in the scaling relations
\begin{eqnarray}
 \label{ZNr}
   &\varepsilon_{\s F}&=\mu~ \propto \xi_I^{-z} ~, \\
   &\xi_I& \equiv 1/k_{\s F}~ \propto \mu^{-\nu^\prime}~,\\
 \label{NuNr}
  &\chi_\circ& ~ \propto \mu^{-\gamma}~,
  \label{GamNr}
\end{eqnarray}
we deduce the critical indices:
\begin{equation}
\label{znuNr}
  \nu^\prime = \frac12~, ~~z=2~, ~~\beta =d  \nu^\prime~,~~\gamma=1-d  \nu^\prime~.
\end{equation}
The ground-state energy density $\mathcal{E}_\circ$ and the density of the grand canonical potential (which is just negative pressure) are
known \cite{LandauV5}:
\begin{equation}
\label{PE0}
  P=\frac32 \mathcal{E}_\circ \propto k_{\s F}^{d+2} ~.
\end{equation}
At zero temperature the critical index of the heat capacity $\alpha$ is usually identified from the scaling of the ground state
thermodynamic potential, see, e.g. \cite{Imada:1998,Fradkin:2013}
\begin{equation}
\label{alphaNr}
  P \propto \mu^{2-\alpha} ~\longrightarrow~ \alpha =1-d  \nu^\prime~.
\end{equation}
The above indices satisfy the hyperscaling relation  $2-\alpha =(d+z) \nu^\prime$
and the Fisher identity \eqref{FisherSc}.

The Lindhard function \eqref{chiQ} for $d$-dimensional non-relativistic fermions is known from the literature.
With the notation $x \equiv q/2k_{\s F}$ the explicit formula for $d=3$ \cite{Girvin:2019}:
\begin{equation}
\label{L3}
 d=3:~~ L(x)= \frac12 + \frac{1-x^2}{4x}\ln \Big|  \frac{1+x}{1-x}  \Big|~.
\end{equation}
In two dimensions, see, e.g., \cite{Chitov:2001}:
\begin{equation}
\label{L2}
  d=2:~~ L(x)=
   \left\{
  \begin{array}{lr}
  1~, ~ x<1 \\ [0.2cm]
  1-\sqrt{1-{1 \over x^2}}~,~x>1 \\
  \end{array}
  \right.
\end{equation}
And for $d=1$:
\begin{equation}
\label{L1}
 d=1:~~ L(x)= \frac{1}{2x}\ln \Big|  \frac{1+x}{1-x}  \Big|~.
\end{equation}

We will also need the correlation function $\mathcal{ D}(\mathbf{r})$ of the spatial density fluctuations
$\Delta n(\mathbf{r}) \equiv n(\mathbf{r})-\bar n$. The average of the square of the Fourier components of the
density fluctuations $\Delta n(\mathbf{q})$ can be obtained from the density response via the fluctuation-dissipation theorem,
but it is easier to calculate it directly from geometrical considerations \cite{LandauV5}:
\begin{equation}
\label{DFl}
  \langle |\Delta n(\mathbf{q})|^2 \rangle = \bar n -  \int \frac{d \mathbf{k}}{(2\pi)^d }
  n_{\s F} (\mathbf{k}) n_{\s F} (\mathbf{k}+\mathbf{q})
  =\bar n -\bar n \Theta(2k_{\s F}-q) S(q)~.
\end{equation}
Here $n_{\s F} (\mathbf{k})= \Theta(-\xi(\mathbf{k}))$ is the Fermi-Dirac distribution function with
$\xi(\mathbf{k}) \equiv \varepsilon(\mathbf{k})- \mu$, and
\begin{equation}
\label{Sq}
   S(x)=
   \left\{
  \begin{array}{lr}
  1-\frac32 x + \frac12 x^3 ~, ~d=3 \\ [0.2cm]
  1- \frac{2}{\pi}( \arcsin x + x \sqrt{1- x^2})~,~d=2 \\[0.2cm]
  1-x~,~d=1 \\[0.2cm]
  \end{array}
  \right.
\end{equation}
Compare to the fluctuations of a classical free field $\varphi(\mathbf{r})$:
\begin{equation}
\label{DFlclass}
  \langle |\Delta \varphi(\mathbf{q})|^2 \rangle \propto \frac{1}{\xi^{-2}+k^2}~.
\end{equation}
The density-density correlation function which is the Fourier transform of \eqref{DFl} reads \cite{LandauV5}:
\begin{equation}
\label{Dr}
 \mathcal{ D}(\mathbf{r}_1-\mathbf{r}_2) \equiv \mathcal{ D}(\mathbf{r})=\bar n \delta(\mathbf{r})-(G(\mathbf{r}))^2~.
\end{equation}
The first term on the r.h.s. of \eqref{DFl} gives the rigid delta-function contribution to $\mathcal{ D}(\mathbf{r})$, while
the fluctuation contribution $\propto S(q)$ combines into square of the spatial representation of a properly taken zero-time limit of
the single-particle Green's function at $T=0$:
\begin{equation}
\label{Gr}
G(\mathbf{r}) =
\int \frac{d \mathbf{k}}{(2\pi)^d }  e^{i\mathbf{kr}} n_{\s F} (\mathbf{k}).
\end{equation}
Explicitly:
\begin{equation}
\label{GNg}
G(\mathbf{r}) = \bar n g(\rho), ~~ \rho \equiv k_{\s F}r~, ~~
\end{equation}
where
\begin{equation}
\label{gd}
   g(\rho)=
   \left\{
  \begin{array}{lr}
  \frac{3}{\rho^3} (\sin \rho - \rho \cos \rho) ~, ~d=3 \\ [0.2cm]
  \frac{2}{\rho} J_1(\rho)~,~d=2 \\[0.2cm]
  \frac{\sin \rho}{\rho}~,~d=1~, \\[0.2cm]
  \end{array}
  \right.
\end{equation}
and  $J_1(\rho)$ is the Bessel function. The asymptotes of the  correlation function are:
\begin{eqnarray}
 \label{glarge}
   \rho \gg 1:~~ &g(\rho)&~ \sim \frac{\cos(\rho + \varphi_d)}{\rho^{(d+1)/2}} \\
   \rho \ll 1:~~ &g(\rho)&~ = 1- \mathcal{O}(\rho^2)
 \label{gsmall}
\end{eqnarray}

We see from the above equations that the density response has a scaling form:
\begin{equation}
\label{chiSc}
  \chi(q) \propto k_{\s F}^{d-2} L \Big( \frac{q}{2k_{\s F}} \Big)~,
\end{equation}
and according to \eqref{DFl} the fluctuations are bound to the region $q<2k_{\s F}$.
Near the critical point
\begin{equation}
\label{chiSc2}
  \chi(q \ll k_{\s F}) \propto k_{\s F}^{d-2} \propto \xi_I^{2-\eta^\prime}~,
\end{equation}
whence
\begin{equation}
\label{eta}
 \eta^\prime=d
\end{equation}
follows. This index satisfies the scaling relation
\begin{equation}
\label{etagamma}
 \gamma = (2-\eta^\prime)\nu^\prime~.
\end{equation}
As an additional consistency check we look at the density correlation function near the critical point
($k_{\s F} \to 0$,~$\xi_I \to \infty$). As follows from the above formulas the behavior of the leading term
\begin{equation}
\label{Deta}
 r \ll \xi_I:~~ \mathcal{D}(r) \propto k_{\s F}^{2d} (1-\mathcal{O}( (r/\xi_I)^2) \sim 1/\xi_I^{(d-2+z+\eta^\prime)}~
\end{equation}
is in agreement with \eqref{eta}. Note that the above indices satisfy another hypercaling relation
$2 \beta = (d+z-2+\eta^\prime) \nu^\prime$.

\subsection{Free Dirac fermions}\label{AppDir}
Using methods of the finite-temperature quantum field theory \cite{Kapusta:2006}, the partition function
of the gas of relativistic Dirac fermions can be expressed as a determinant of the inverse Green's function
operators
\begin{equation}
\label{GrD}
  \hat G^{-1}(\mathbf{k},\omega_n)=\hat \gamma_0(i \omega_n +\mu)+  \mbox{\boldmath$\hat \gamma$}  \mathbf{k} -m~,
\end{equation}
which include in addition to the momentum-(Matsubara) frequency space, the Dirac $\hat \gamma$-matrices.
The condition for zeros of the partition function amounts to
\begin{equation}
\label{LYZD}
  \prod_{\mathbf{k},n} \big[(i \omega_n +\mu)^2-\varepsilon^2(\mathbf{\mathbf{k}})\big]=0,
\end{equation}
where $\varepsilon(\mathbf{\mathbf{k}})=\pm \sqrt{\mathbf{k}^2+m^2}$ is the spectrum of the Dirac fermions. We
consider $\mu \geq 0$ which accounts over the surplus $\bar n$ of fermions ($\sqrt{\mathbf{k}^2+m^2}$) over antifermions
($-\sqrt{\mathbf{k}^2+m^2}$). The negative band is filled at $\mu \geq 0$. For simplicity we will present the results per one
fermionic flavor (spin). The critical properties of fermions with the Dirac spectrum are relevant not only for relativistic electron gas
\textit{per se}, but also in the context of condensed matter physics \cite{Volovik:2003,Shen:2017}.

For the Dirac fermions $\mu(T=0)= \varepsilon_{\s F} =\sqrt{k_{\s F}^2+m^2}$ and $\mu_c=m$. Similarly to non-relativistic fermions,
no phase transition can occur at $T \neq 0$, since $\nexists~ k \equiv |\mathbf{k}| \in  \mathbb{R}$ to solve \eqref{LYZD}.
At $T=0$ in the notations  \eqref{qkap} we find
\begin{equation}
\label{QKapDir}
  \mu >m:~~
   \left\{
  \begin{array}{lr}
  q=k_{\s F}=1/\xi_I=\sqrt{\mu^2-m^2} ~\rightarrow~ \nu^\prime =1/2 \\ [0.2cm]
  \kappa=0~~(\xi= \infty) \\
  \end{array}
  \right.
\end{equation}
The system is gapless with the linear excitation spectrum $E(k) \approx  |k-k_{\s F}| k_{\s F} /\varepsilon_{\s F}$ near the FS.
The characteristic energy scale  (depth of the Fermi sea) $\varepsilon_{\s F} -\varepsilon(0) =\mu-m$ implies $z\nu^\prime=1~\rightarrow~z=2$.
At the QCP $\mu =m$ the spectrum is $E(k)=k^2/2m + \mathcal{O}(k^4)$, also in agreement with $z=2$.
The order parameter $\mathcal{P}$ and $\bar n$ are given by the formulas (\ref{OPnr},\ref{avn}) with $k_{\s F}$ from \eqref{QKapDir},
while
\begin{equation}
\label{chi0Dir}
\frac{\partial \bar n}{\partial \mu}=\frac{d \Omega_d}{(2 \pi)^d} \mu k_{\s F}^{d-2}~.
\end{equation}
The ground-state pressure \cite{LandauV5}
\begin{equation}
\label{PDir}
  P= \frac{\Omega_d}{(2 \pi)^d}  \int_0^{ k_{\s F}} dk k^{d-1} (\mu  - \sqrt{k^2+m^2}) \propto (\mu-m)^{1+d/2} ~.
\end{equation}
(Note that the ground-state energy density scales differently: $\mathcal{E}_\circ \propto (\mu-m)^{d/2}$).
At this point one can straightforwardly find all critical indices which are the same as for the non-relativistic fermions found in Sec.~\ref{AppNR}.
In particular, the value of index $\eta^\prime$ \eqref{eta} can be inferred directly from \eqref{etagamma}. Getting $\eta^\prime$ directly from the response and correlation functions amounts to the calculation of the matter contribution to the polarization operator $\Pi(\mathbf{q})$ in lowest order \cite{Kapusta:2006}, which is straightforward, but tedious. Luckily, it is not necessary, since the limiting behavior of the correlation functions we need to know to extract $\eta^\prime$, corresponds to the nonrelativistic limit $ k_{\s F} \ll m$, when the results of Sec.~\ref{AppNR} are applicable.\footnote{\label{Kapusta} For instance,
one can check by direct inspection that the matter contribution of the fermionic loop to $\Pi(\mathbf{q})$ given in Ch.~6 of \cite{Kapusta:2006} yields the series
identical (in the limit $ k_{\s F} \ll m$) to the expansion of $L(x)$ defined in \eqref{L3}.}

In the gapped phase
\begin{equation}
\label{Dmet}
  0< \mu <m:~~
   \left\{
  \begin{array}{lr}
  q=0  \\ [0.2cm]
  \kappa= \sqrt{m^2-\mu^2}~\rightarrow~ \nu =1/2 \\
  \end{array}
  \right.
\end{equation}
The linear gap dependence $\Delta=m-\mu$, i.e., $z \nu=1$, implies that $z=2$ on the both sides from the critical point. We also find $\nu=\nu^\prime$
which allows us to replace $\nu^\prime \mapsto \nu$ in the results for the critical indices and thus to conclude that all
the scaling relations we found, are the standard ones known from the theory of critical phenomena. Note however that the equality
$\nu=\nu^\prime$ does hold in general.

At $\mu<m$ we get
an incompressible state, since $\chi_\circ=0$, the counterpart of the plateau states in chains and ladders.

\end{appendix}

\bibliography{C:/Users/gchitov/Documents/Papers/BibRef/CondMattRefs}
%
%
\end{document}